\documentclass[twocolumn]{openjournal}

\usepackage[utf8]{inputenc}
\usepackage[T1]{fontenc}
\usepackage{float}
\usepackage{xcolor}
\usepackage{ulem}
\usepackage{graphicx}	% Including figure files
\usepackage{amsmath}	% Advanced maths commands
\usepackage{comment}
\usepackage{color}
\usepackage{hyperref}
\hypersetup{colorlinks=true,
	urlcolor=blue,  % color of external links
	linkcolor=blue,  % color of toc, list of figs etc.
	citecolor=blue, % color of links to bibliography
    menucolor=blue, % color for Acrobat menu buttons
    urlcolor=blue}  % color for \url{...} links

\def\mpc{\, {\rm Mpc}}

\def\hkpc{h^{-1}\, {\rm kpc}}

\def\msun{\, M_{\odot}}

\def\simlt{\lower.5ex\hbox{$\; \buildrel < \over \sim \;$}}
\def\simgt{\lower.5ex\hbox{$\; \buildrel > \over \sim \;$}}

\newcommand{\review}[1]{\textcolor{black}{#1}}

\begin{document}

% Title of the paper, and the short title which is used in the headers.
% Keep the title short and informative.
\title{The evolution of galaxy morphology from redshift $z=6$ to $3$: Mock JWST observations of galaxies in the ASTRID simulation}

% The list of authors, and the short list which is used in the headers.
% If you need two or more lines of authors, add an extra line using \newauthor
\author{Patrick LaChance$^{1,*}$}
\author{Rupert Croft$^{1,2}$}
\author{Yueying Ni$^{3}$}
\author{Nianyi Chen$^{1}$}
\author{Tiziana Di Matteo$^{1,2}$}
\author{Simeon Bird$^{4}$}
\thanks{$^*$E-mail:plachance@cmu.edu}
% List of institutions
\affiliation{$^{1}$ McWilliams Center for Cosmology, Department of Physics, Carnegie Mellon University, Pittsburgh, PA 15213 USA \\
$^{2}$ NSF AI Planning Institute for Physics of the Future, Carnegie Mellon University, Pittsburgh, PA 15213, USA \\
$^{3}$ Harvard-Smithsonian Center for Astrophysics, Harvard University, 60 Garden Street, Cambridge, MA 02138, USA \\
$^{4}$ Department of Physics \& Astronomy, University of California, Riverside, 900 University Ave., Riverside, CA 92521, USA}

% Abstract of the paper
\begin{abstract}
We present mock JWST observations for more than \review{250,000} different galaxies from the \texttt{Astrid} simulation with $3 \leq z \leq 6$. The mock observations are made using the BPASS stellar SED model, and a simple dust model. They are then viewed through NIRCam filters, convolved with a PSF, have noise added, and are drizzled together to emulate the Cosmic Evolution Early Release Science (CEERS) survey. We analyse this dataset by computing a number of morphological measures and find our catalog to have comparable statistics to similar mock catalogs, and the first release of CEERS data. We find that most of the Sersic indices of galaxies in our redshift range are lower than observed, with most having $n \leq 1$. Additionally, we observe the sizes of galaxies of all masses to increase from redshift $z=6$ to redshift $z=3$ consistent with other results. The number of galaxies in our catalog allows us to examine how relationships like the mass-size relation evolve with redshift, and compare the accuracy of a variety of traditional galaxy classification techniques (Sersic fit, Asymmetry-Concentration, and Gini-$M_{20}$) within our redshift range. We find the mass-size relation to be nearly flat at redshift $z=6$, and consistently increases as redshift decreases, and find the galaxy classification methods have minimal correlation with each other in our redshift range. We also investigate the impact that different stages of our imaging pipeline have on these morphological measures to determine how robust mock catalogs are to different choices at each step. Finally, we test the addition of incorporating light from AGNs into our pipeline and find that while the population of galaxies that have significant AGN luminosity is low, those galaxies do tend to have higher Sersic indices once the AGN luminosity is added, rectifying some of the systematic bias towards lower Sersic indices present in our dataset.
\end{abstract}

% Select between one and six entries from the list of approved keywords.
% Don't make up new ones.
\keywords{Galaxy evolution, Galaxies, Surveys, Hydrodynamical Simulations, High Redshift}

\maketitle

%%%%%%%%%%%%%%%%%%%%%%%%%%%%%%%%%%%%%%%%%%%%%%%%%%

%%%%%%%%%%%%%%%%% BODY OF PAPER %%%%%%%%%%%%%%%%%%

\section{Introduction}

\begin{figure}
    \centering
    \includegraphics[width=1.0\columnwidth]{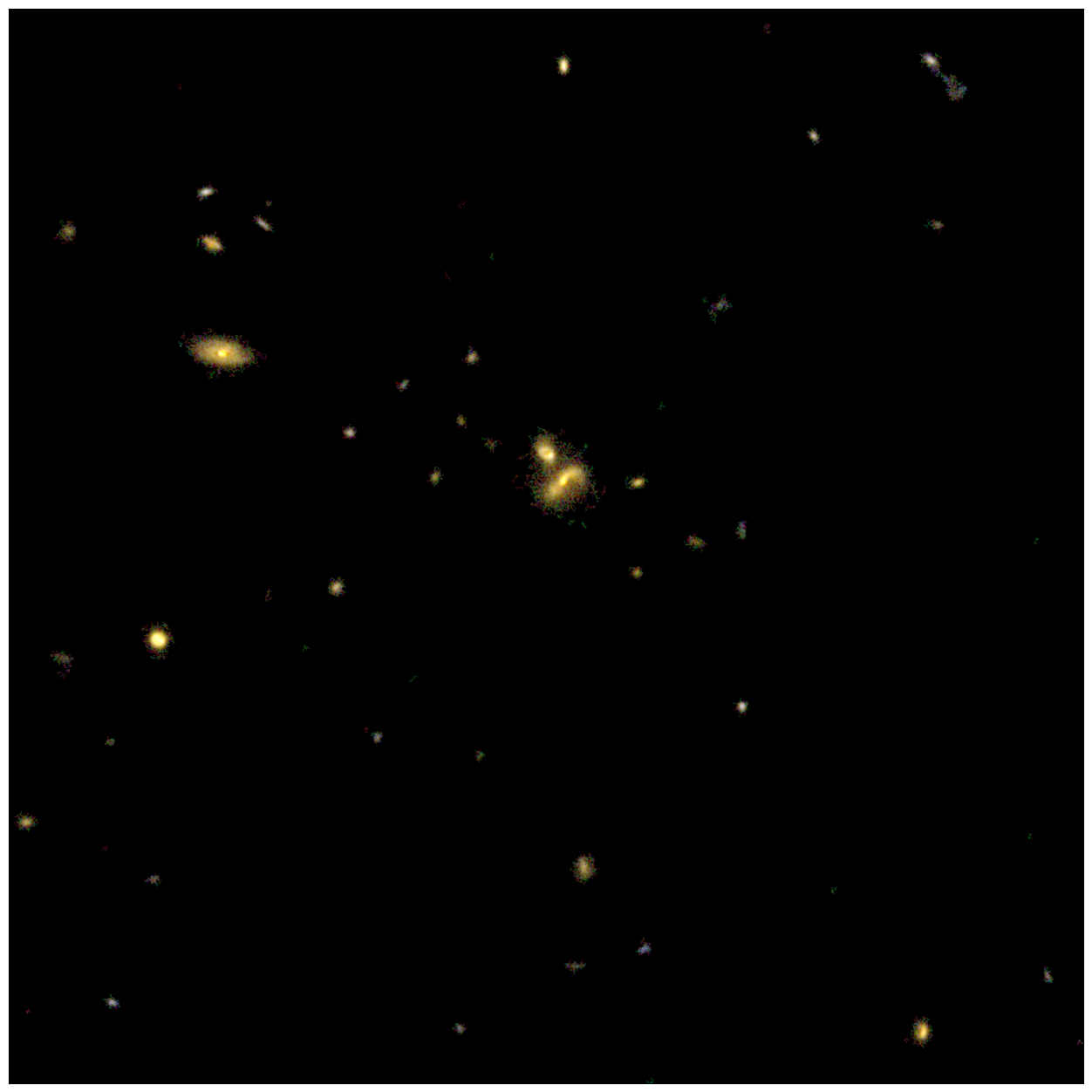}
    \caption{A false color mock observation of a group of galaxies in the Astrid simulation using JWST NIRCAM filters F356W, F277W and F150W. Created by drizzling the individual observations for each galaxy in the group onto a single observation. The image is $\sim 46.7"$ on a side, and includes $\sim 50$ galaxies.}
    \label{fig:Group_image}
\end{figure}

Our understanding of galaxies in the local universe is substantial, but there are still many gaps in our understanding of how the Hubble Sequence we see in the local universe came to be. Many high quality galaxy observations of nearby galaxies provide a very detailed picture of the galaxy population at redshift 0. Catalog-based research such as Galaxy Zoo \citep{Lintott_2011} has given us insight into many aspects of galaxies in the local universe confirming that spiral galaxies tend to be forming more stars, and be bluer while ellipticals are quiescent and redder \citep{Lintott_2008, Cappellari_2016}, but also allowing deeper analysis separating the impact of environment from morphology \citep{Bamford_2009, Skibba_2009}, and even understanding what is different about galaxies that break these established trends, such as red spirals. \citep{Masters_2010_a, Masters_2010_b}

 As redshift increases the volume of available data on galaxy populations decreases because of the observational challenges of acquiring high resolution observations of galaxies at greater distances. Despite those challenges there are catalogs of galaxies at higher redshifts that show which aspects of the galaxy population have stayed relatively consistent throughout cosmic history, and which have evolved with time. Observations of more distant galaxies show the proportion of disk, elliptical, and irregular galaxies evolving, with irregular galaxies increasing in prevalence as higher redshifts \citep{Abraham_1996, Giavalisco_1996, Conselice_2000, Lotz_2006, Ravindranath_2006, Driver_1995, Papovich_2005, Mortlock_2013, Huertas_Company_2016}. Additionally, the sizes of galaxies have grown significantly since cosmic noon \citep{vanderwel_14, Daddi_2005, Trujillo_2007, Grogin_2011, Mowla_2019_b, Koekemoer_2011, Buitrago_2008}.

Despite those shifts in the morphology prevalence, and galaxy size, there is a body of work showing that the properties of the different morphological classes stay relatively consistent, with most star-forming galaxies being disks, and most quiescent galaxies being spherical or compact \citep{Wuyts_2011, Lee_2013, Mortlock_2013, Zhang_2019, Kartaltepe_2015}.

Further understanding what causes some of these properties to evolve, and the others to remain unchanged is vital to expanding our understanding of how galaxies form and evolve throughout cosmic history. There are many big questions that remain unanswered, including how frequent and impactful mergers are for the growth of galaxies compared to more steady accretion, and whether galaxies tend to evolve inside-out or outside-in. New state of the art observatories such as the James Webb Space Telescope (JWST) have the capability to increase the amount and quality of data about high redshift galaxies by orders of magnitude, and will play a key role in answering these questions. Early results are already being published that show the potential for a revolution in our knowledge of galaxies at redshifts $z=3$ and above. These include population analyses like \citet{CEERS_1, CEERS_3}, new conclusions about galaxy sizes \citep{galaxies_cosmic_noon}, and morphologies \citep{Panic_discs, Red_spirals}, insight into the dust characteristics of high redshift galaxies \citep{Kirkpatrick_2023, LeBail_2023}, and new insights into the characteristics of Active Galactic Nuclei (AGNs) at these high redshifts \citep{Barro_2023, Yang_2023}.

Mock observations from simulations can augment the results of observatories in advancing this field. They can provide valuable calibration datasets to compare with the morphology measurements made from observations, draw direct connections between certain observational features, and physical properties of galaxies within the simulation, and provide insight into rare objects that are unlikely to be found via observation.

We create our mock catalog from the \texttt{Astrid} simulation \citep{Astrid_BHs, Astrid_galaxy_formation} and make our mock observations to emulate the characteristics of the NIRCam instrument on JWST. Specifically we try to mimic the observation strategy used by the Cosmic Evolution Early Release Science (CEERS) survey \citep{CEERS_proposal, CEERS_epoch_1_strategy}.

In section \ref{sec:methods} we describe the mock observation creation pipeline we developed. In section \ref{sec:results} we discuss the results of performing morphological fits on our galaxy dataset, and compare our results to those of similar datasets. In section \ref{sec:analysis} we examine the relationships between these fit parameters, how they evolve with time, and what those relationships indicate about the galaxy population at these redshifts. Finally, in section \ref{sec:Summary} we summarize our findings throughout this work, discuss how those results are connected, and possible future work in this area.

\section{Methods}
\label{sec:methods}
In this work we create an imaging pipeline that processes the galaxies present in the \texttt{Astrid} simulation into NIRCAM-like observations. We follow a similar, but slightly simplified pipeline compared to \citet{TNG50_CEERS} in order to process the large number of galaxies present in Astrid in a timely fashion. 

\begin{figure}
    \centering
    \includegraphics[width=\columnwidth]{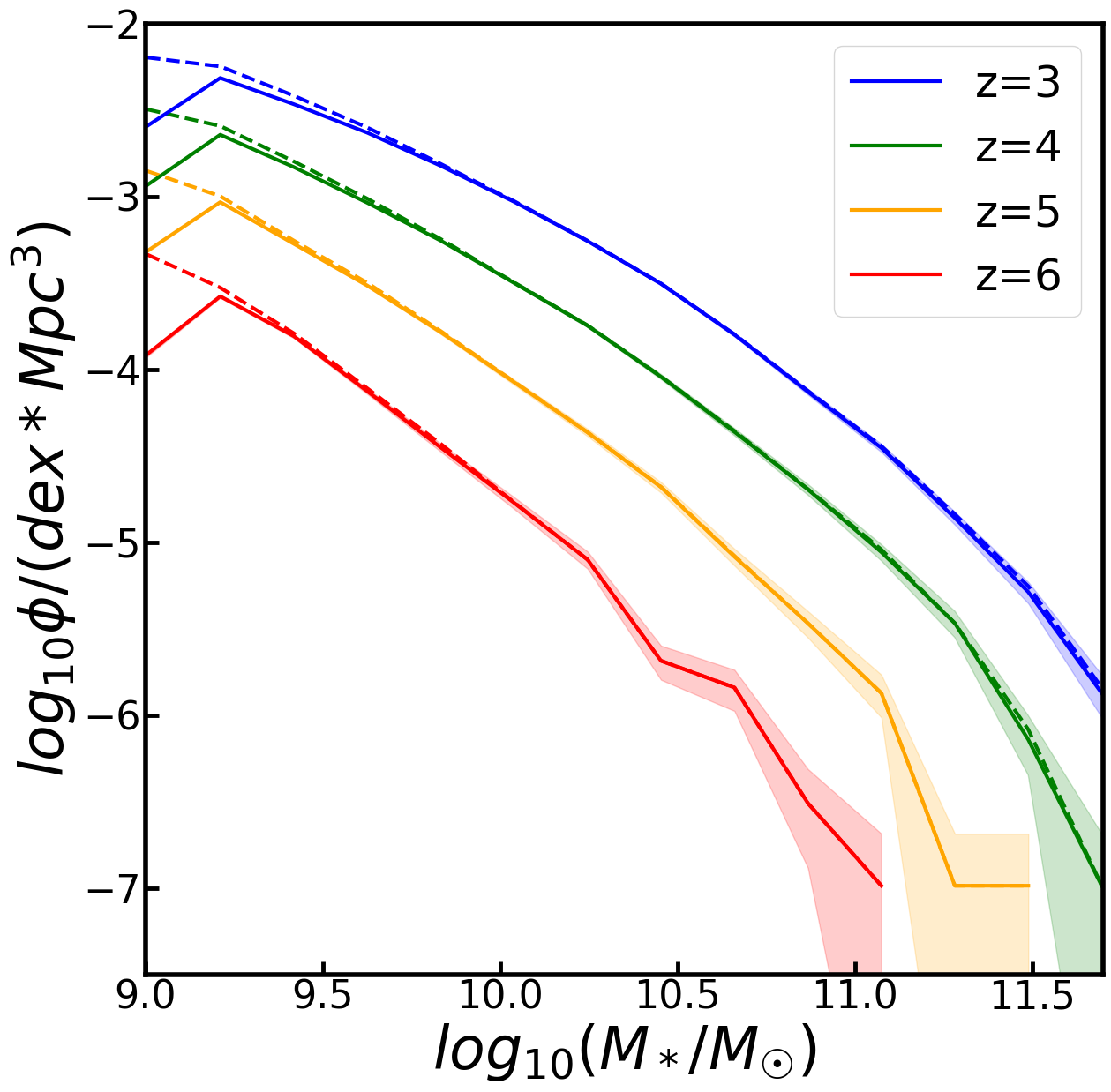}
    \hfill
    \includegraphics[width=\columnwidth]{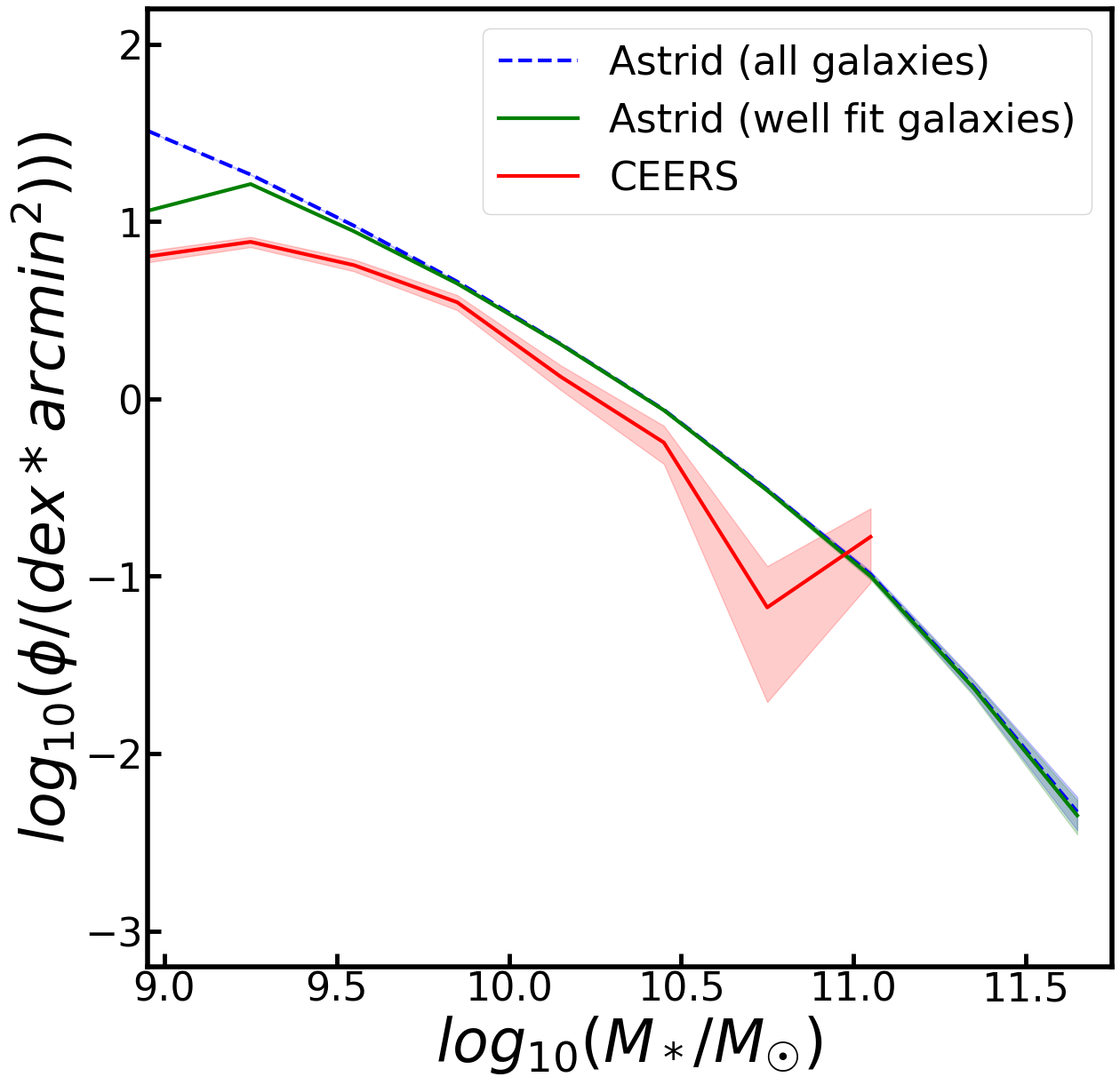}
    \caption{The stellar mass function of our galaxy sample, and the CEERS sample. The top panel shows the galaxy mass function for each of the 4 snapshots present in our sample, with the solid line indicating the galaxies that were detected and morphologically fit without issue, and the dashed line shows the overall subhalo mass function regardless of fitting. The bottom panel shows a survey area-scaled galaxy mass function for our dataset, and the CEERS dataset. In both panels the Poisson error is indicated by the shaded region around each line.}
    \label{fig:gal_sample}
\end{figure}

\begin{figure*}
    \centering
    \includegraphics[width=2.0\columnwidth]{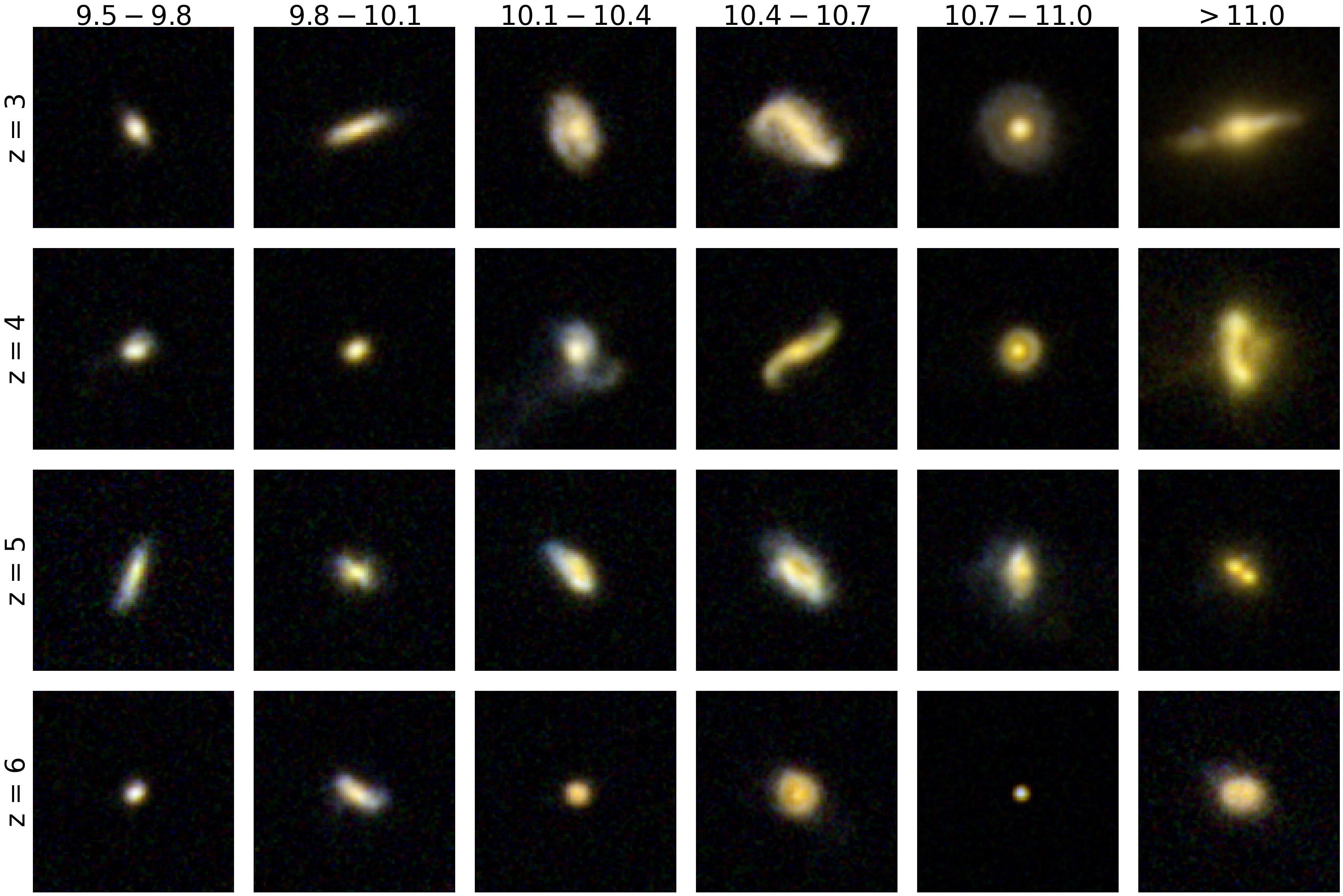}
    \caption{Example false color images of galaxies from all 4 redshift samples, across a range of stellar masses. Filters F356W, F277W, and F150W were used for the red, green and blue channels respectively. Each image is 3" on a side. The number ranges above each column indicate the range of $log_{10}(M_*/\msun)$ for galaxies shown in that column}
    \label{fig:galaxy_grid}
\end{figure*}

\subsection{The Astrid Simulation}
\label{sec:sim}

\texttt{Astrid}  \citep{Astrid_BHs, Astrid_galaxy_formation} is a cosmological simulation run with the \texttt{MP-Gadget} a version of the cosmological simulation code \texttt{Gadget-3} \citep{Springel_2005}. It has a box of side-length $250 \mpc h^{-1}$ and initially contains $2\times 5500^3$ particles (half dark matter, half SPH gas particles). This gives it a volume larger than the Illustris TNG300 simulation, but with a mass resolution similar to the TNG100 and EAGLE simulations (Illustris TNG: \citealt{Springel_2018, Nelson_2018, Marinacci_2018, Naiman_2018}; EAGLE: \citealt{Schaye_2015}). The cosmological parameters employed for \texttt{Astrid} are based on the results of \citet{Planck_2020} ($\Omega_0 = 0.03089$, $\Omega_{\Lambda} = 0.6911$, $\Omega_b = 0.00486$, $\sigma_8 = 0.82$, $h = 0.6774$, $n_s = 0.9667$). \texttt{Astrid} has recently reached $z=1.0$.

The star formation model in Astrid is the same as was used by \citet{Feng_2016} for the \texttt{BlueTides} simulation, and is based on the star formation model described in \citet{Springel_2003}. The gas cooling model which effects the star formation process includes both radiative and metal-line cooling. The treatment of gas particle cooling includes the self-shielding of dense gas outlined in \citet{Rahmati_2013} and the star formation criteria includes a correction to account for the formation of molecular hydrogen \citep{Krumholz_2011}. This process produces star particles that are 1/4 the mass of the gas particle they are formed from. 

These gas and star particles along with the dark matter, and black hole particles are collected into halos using a friends of friends algorithm \citep{Davis_1985} and then post processed into subhalos using the software \texttt{SUBFIND} \citep{Springel_2001}. These subhalos represent the halo substructures associated with galaxies within the simulation, which we use as the basis for producing our galaxy images. General analysis of these subhalos was performed in \citet{Astrid_galaxy_formation} which found good agreement between the UV luminosity function and Galaxy Stellar Mass Function of \texttt{Astrid} and observational data in the redshift range we examine in this work ($3 \leq z \leq 6$).

\subsection{Galaxy Sample}
\label{sec:sample}

\review{Using the subfind catalogs for the simulation snapshots at redshifts $z=3$, 4, 5, and 6, we include all subhalos with a subfind stellar mass above $log_{10}(M_*/\msun) = 9$}. We pass those subhalos to the mock observation pipeline outlined in section \ref{sec:pipeline}, and perform morphological fitting on those mock observations as outlined in section \ref{sec:morph}. That fitting process produces flags for any observations that have issues. Given that this is a mock catalog, these issues are largely limited to observations with low signal-to-noise ratios, sources that are difficult to segment, and \review{irregular galaxies} that are difficult to fit. \review{We removed any galaxies that had their basic fitting or Sersic fit flagged in any way, and all galaxies with a signal-to-noise ratio below 2.5.} We discuss the impact of removing these poorly fit galaxies below and in Section \ref{sec:image_stages}.

The stellar mass function for our galaxy sample both before and after removing the poorly fit galaxies can be seen in figure \ref{fig:gal_sample}. \review{For this analysis, and all future mass-dependent analysis we recalculate the stellar mass of all galaxies to be the stellar mass within twice the half-mass radius in order to make it more analogous to observational measurements of stellar mass based on an aperture. The population of all subhalos above $log_{10}(M_*/\msun) = 9$ in the simulation is quite similar to the subset of that population that is able to be detected and well fit. That said, there is a noticable difference at the low mass end, as the galaxies in this regime are the most likely to encounter issues with low signal-to-noise ratios, segmentation, and fitting.}

In order to make a more direct comparison to the CEERS catalog we make a subsample of our population that is intended to represent a very large survey that could be taken by making use of the volume of the Astrid simulation between the redshift values represented by each snapshot  (the $z=3$ snapshot is used for $3 \leq z < 3.5$, the $z=4$ snapshot for $3.5\leq z <4.5$ and so on). We choose a survey area of $\sim 15,000$ square arcmins as the largest comoving volume that requires at a single redshift is slightly smaller than the Astrid volume. The redshift $z=4$ snapshot accounts for the largest comoving volume, as it is the lowest redshift that covers a whole redshift range ($3.5 \leq z \leq 4.5$) rather than a range of half a redshift like the redshift 3 snapshot ($3 \leq z \leq 3.5$).

While this mock survey population is the closest we can get to a real survey with only integer redshift snapshots, it does have some variation from a real survey's redshift population distribution due to using single redshift snapshots to represent redshift ranges. We produce an weighted version of this population that emulates the redshift distribution of the CEERS dataset from z=3 to z=6, and see very minimal impacts on the population statistics we analyze as discussed in \ref{sec:dataset_comparisons}.

Using this "mock survey" we compare our stellar mass function to that of CEERS which can be seen in the lower panel of fig. \ref{fig:gal_sample}. \review{It is clear we have an over abundance of galaxies at all masses compared to CEERS (by a factor of $1.6-1.9$), but when compared to other larger surveys the mass function of Astrid matches observations quite well as seen in \citet{Astrid_galaxy_formation}.} This indicates the overabundance seen here could be the result of further selection effects due to the observation strategy, source matching, and other observational effects that we did not attempt to model. \review{Despite this discrepancy the overall characteristics of our galaxy samples match that of the CEERS survey, and similar surveys and mock catalogs well enough that they can be compared.}

\begin{figure*}
    \centering
    \includegraphics[width=1.0\columnwidth]{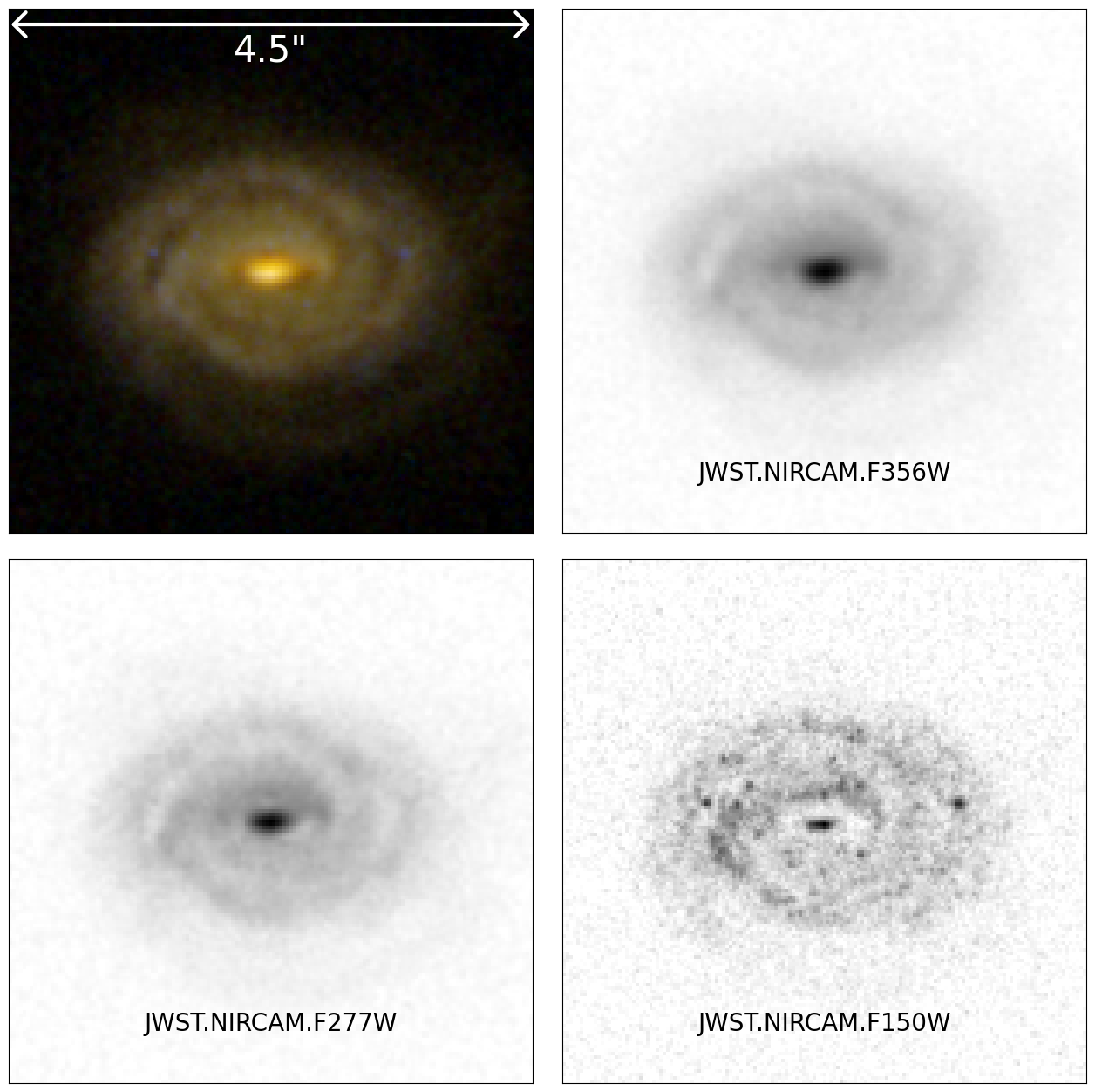}
    \hspace{5mm}
    \includegraphics[width=1.0\columnwidth]{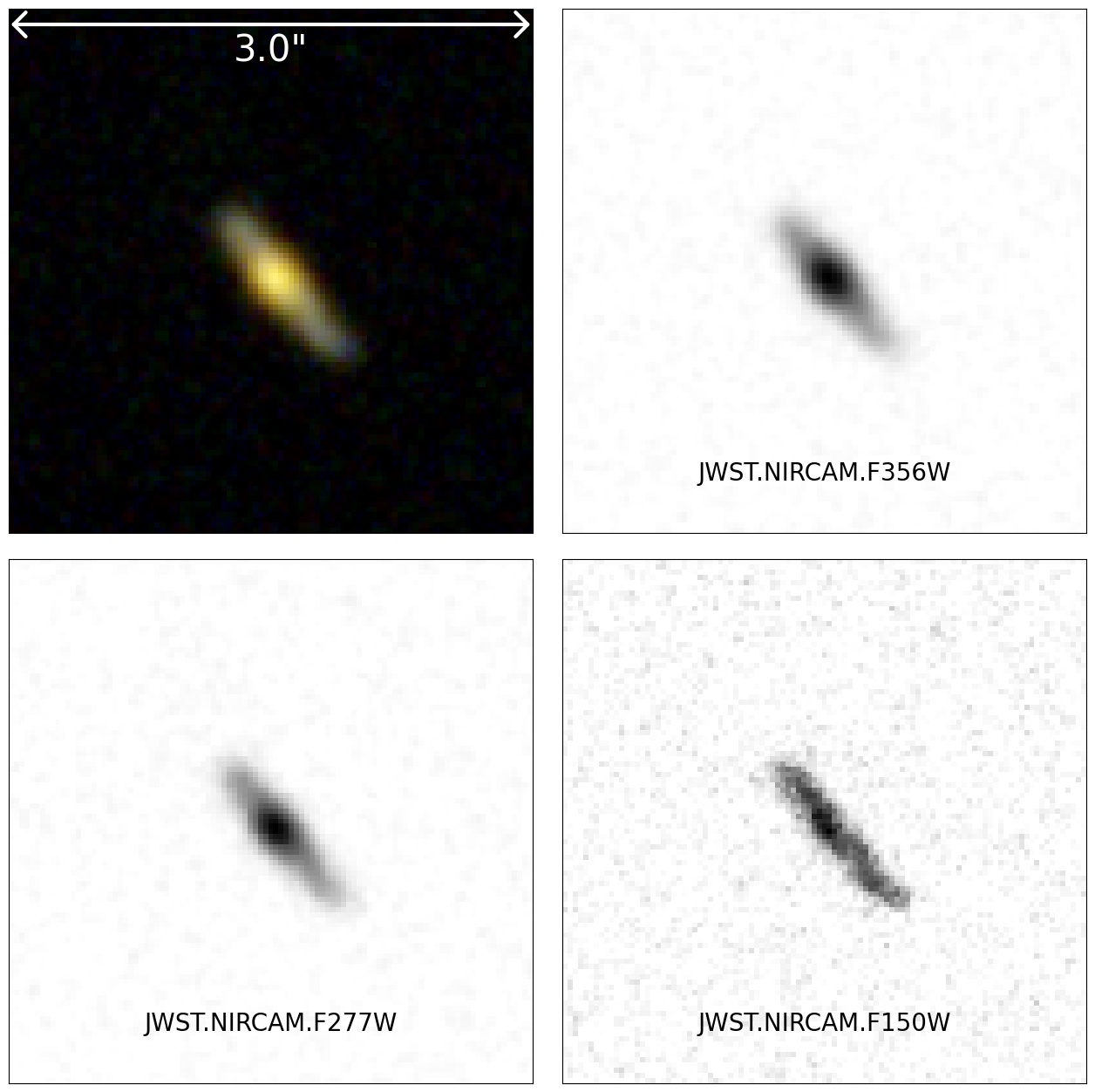}
    \caption{Example false color and individual filter images of two galaxies from the Astrid simulation. The false color images are made by assigning the F356W, F277W, and F150W filter observations to each of the red, green, and blue channels of the image respectively. The galaxy on the left is a $log_{10}(M_* / \msun) = 11.64$ galaxy from the redshift 3 snapshot, and the galaxy on the right is a $log_{10}(M_* / \msun) = 10.39$ galaxy from the redshift 4 snapshot.}
    \label{fig:image_decomp}
\end{figure*}

\subsection{Mock Observation Pipeline}
\label{sec:pipeline}

We create our synthetic observations using the gas and star particles \review{that are} present in each galaxy. The properties of each star particle are used to calculate its spectral energy density (SED) , and the gas is used to make a full three-dimensional dust density map which is used to calculate the amount of dust occlusion the light from each star particle would experience. We pass these dust occluded SEDs through mock NIRCAM filters in order to produce stellar luminosities associated with each NIRCAM observational band. We combine those stellar luminosities with each star particle's position and smoothing length to create noiseless synthetic observations with the resolution of their corresponding NIRCAM filters. Finally, we emulate some instrumentational effects by applying a  NIRCAM-like point-spread function (PSF) to the image, adding Poisson background noise and drizzling multiple observations of each galaxy using the CEERS drizzling strategy. \citep{CEERS_epoch_1_strategy}

\subsubsection{Calculating Stellar Luminosities}
\label{sec:stellar_lums}
In order to calculate the luminosity of each star particle in each of the NIRCAM filters we start by assigning to each of our star particles an SED based on the closest point in a simple stellar population model. The stellar population model we use is the Binary Population and Spectral Population Synthesis model \citep[BPASS version 2.2.1;][]{Stanway2018} which  uses
mass, age, and \review{metallicity} to determine the SED for each star particle.

Next, we calculate the dust attenuation for each star. We use a slightly simplified version of the process detailed in \citet{FLARESII}. Specifically, we find the density of metals along the line of sight, and calculate the optical depth for a given wavelength $\tau_{\lambda}$ as
\begin{equation} \label{eq1}
\tau(\lambda) = -\kappa \Sigma(x,y,z) \left(\frac{\lambda}{0.55 \mu m}\right)^{-\gamma}
\end{equation}
where $\kappa$ is a tuning parameter, $\Sigma(x,y,z)$ is the metal surface density of the star, and $\gamma$ is a scaling factor in the dust model \citep{Wilkins17}. This same procedure was used in \citet{Astrid_galaxy_formation}, where the choice of $\kappa=10^{4.1}$ and $\gamma = -1$ was made by calibrating to the observed galaxy UV-luminosity function at $z=4$. We choose to use these same parameters in order to remain consistent with their work.

We calculate the metal surface density by creating a three dimensional metal density map from the gas particles, weighted by their mass and \review{metallicity}. A 3D Gaussian kernel with $\sigma = 1 \hkpc$ is applied to the whole map in order to approximate smoothing each gas particle by its SPH-kernel. Using this map, we calculate the metal surface density by integrating the metal density between the star and the edge of the galaxy along the line of sight.

This process results in the full spectrum dust extincted SED for each star particle. In order to calculate the brightness of each star in individual NIRCAM filters, we multiplied these SEDs by the transmission curves for each of the filters provided in the JWST user documentation \citep{JWST_user_doc}. Given that the above SED calculations produce the restframe SED, the wavelengths of these transmission curves were properly adjusted to account for the redshift of the galaxy being observed ($\lambda_{\rm{adjusted}}(z) = \lambda*\frac{1}{1+z}$). This final step produces the luminosity density of each star as observed through a number of NIRCAM filters. We used the \texttt{SynthObs} \citep{synthobs_cite} code package in order to carry out the SED assignment, dust attenuation, and filter application portions of this pipeline.

\begin{figure*}
    \centering
    \includegraphics[width=1.0\columnwidth]{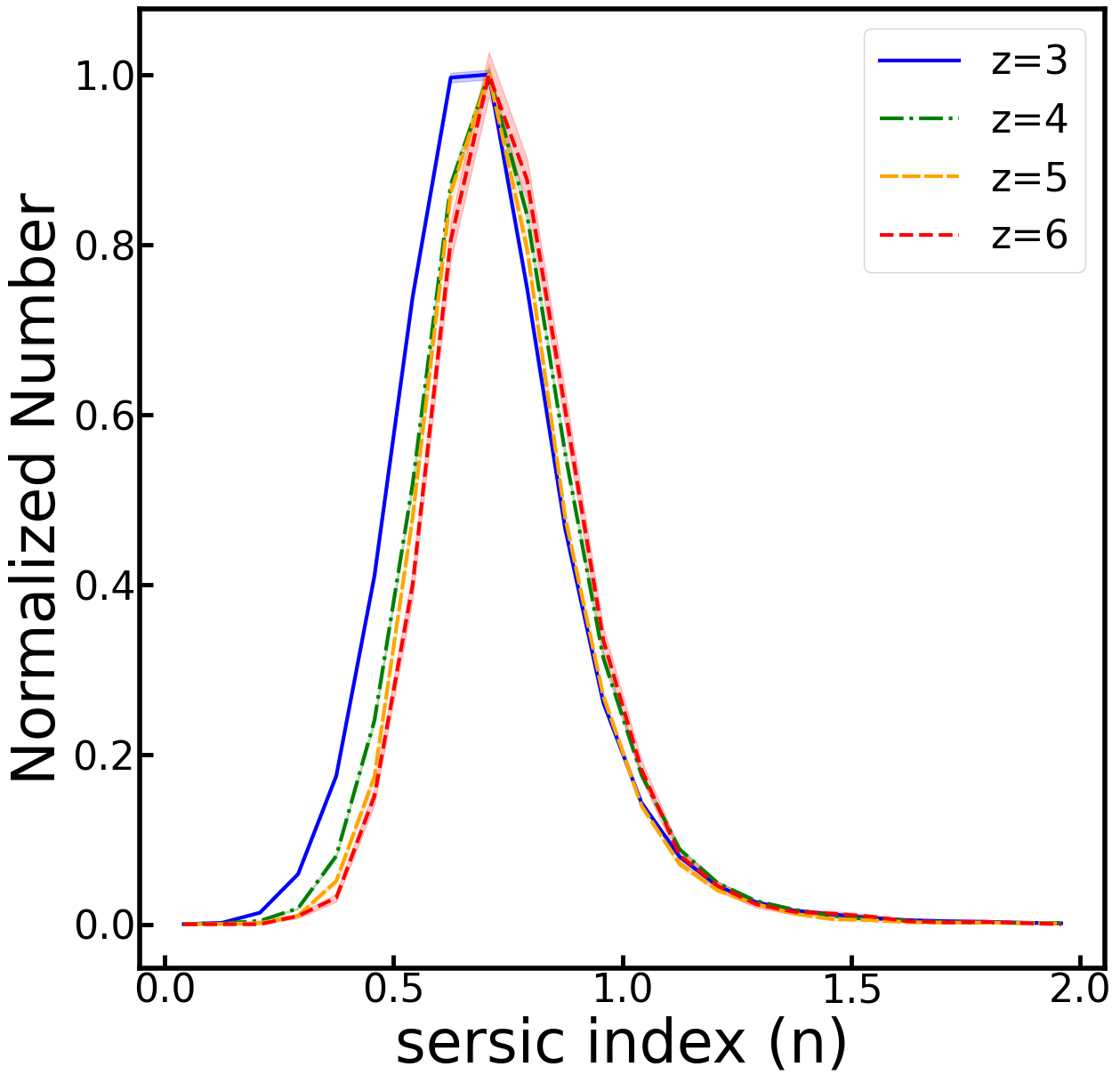}
    \includegraphics[width=1.0\columnwidth]{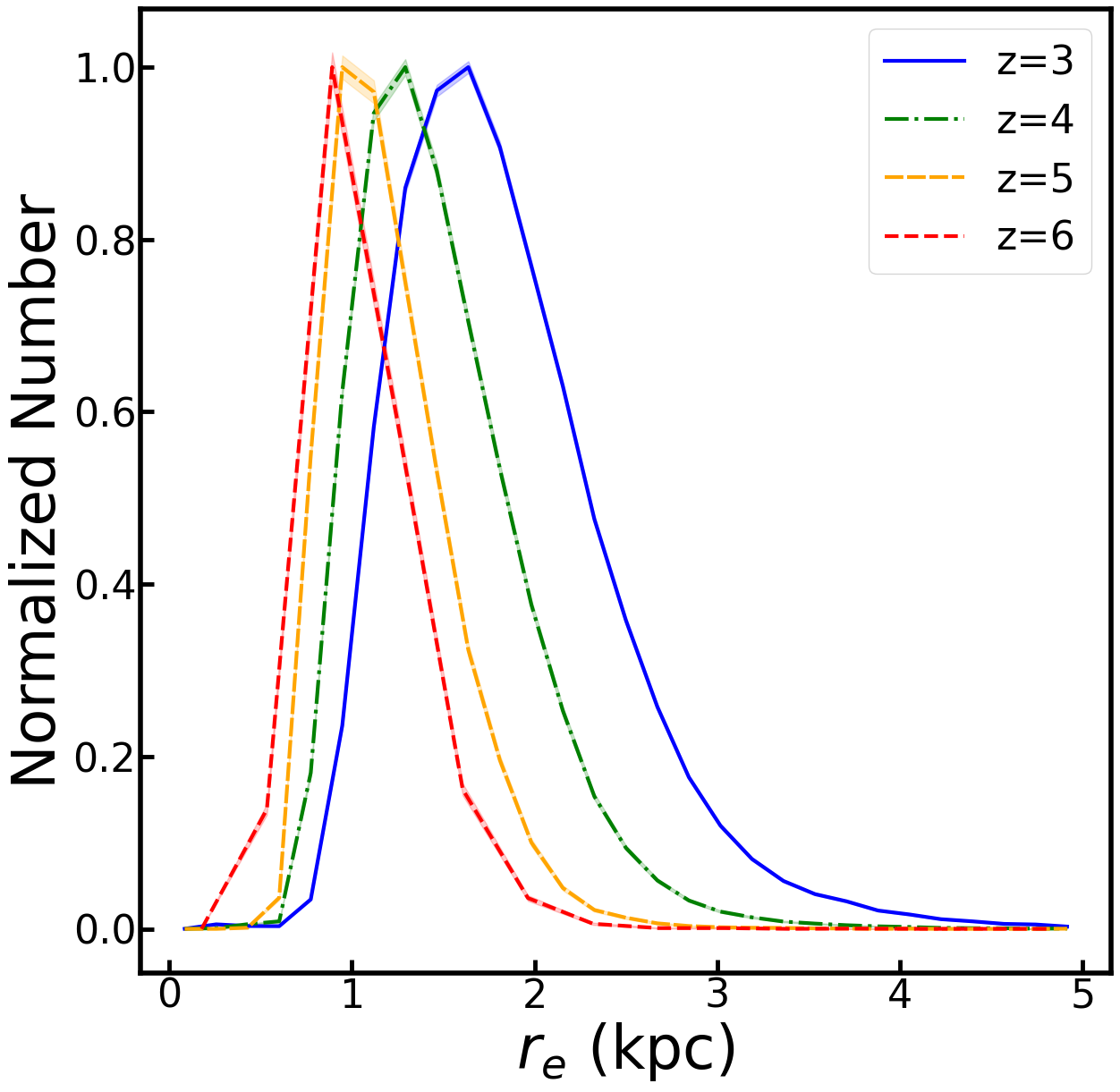}
    \caption{Normalized histograms of the Sersic indicies, and Sersic effective radii of galaxies at each redshift. The histograms for each redshift are normalized so that the bin with the largest 
    number of galaxies has height 1. Poisson error is indicated by the shaded region around each line.}
    \label{fig:Sersic_hist}
\end{figure*}

\subsubsection{Image Creation}
\label{sec:image_creation}
Using the stellar luminosity densities we create the synthetic observations. We begin by creating noiseless synthetic observations using the simulation visualization code \texttt{Gaepsi2}\footnote{Code available at \url{https://github.com/rainwoodman/gaepsi2}.} . We use the intensity of each star particle in the corresponding filter, place it according to position in the galaxy and smooth that intensity according to each particles' smoothing length to make the noiseless images. The smoothing length of each star particle is set to be the distance to the nearest 60 stars, which is used to distribute the stellar light using a 2D SPH kernel. We create these noiseless images with the size and resolution characteristics of the NIRCAM filters they correspond to. Specifically the pixel size of the short and long wavelength images are $0.031"$ and $0.063"$ respectively, and the minimum image size is 100x100 short wavelength pixels. We create 3 images of each galaxy, following a 3-point dither pattern to model the dither pattern used by CEERS. Each image was created with an exposure time of $t=945 s$ to create an overall exposure time of $t=2835 s$ to match the CEERS observation strategy \citep{CEERS_proposal, CEERS_epoch_1_strategy}.

In addition to emulating the size and resolution of NIRCAM observations, we also convolve a PSF and add background noise to our noiseless images to simulate the most dominant observational effects present in NIRCAM images. The PSF we used is a 2D Gaussian distribution with a full width-half max of 2 pixels for both short and long wavelength filters. This corresponds to the reported average FWHM of $0.062$ and $0.126$ arcsec for the short and long wavelength bands.

\review{We used a constant approximation of the background radiance of $0.25$ 
$\rm MJy/sr$ across all filters, which is added to all observations. After the background is added, we simulate the photon noise of the observation using uncorrelated Poisson noise on the detection counts in each pixel. We calculate the expected electron counts in each pixel ($N_{pix,exp}$) using the following formula,
\begin{equation}
N_{\rm pix,exp} = \frac{L_{\rm pix}}{{\rm sens}_{\rm avg}} *\rm e_{\rm gain}* t,
\end{equation}
where $\rm sens_{\rm avg}$ is the average pixel sensitivity in $\rm (MJy/sr) / (DN/s)$, $L_{\rm pix}$ is the radiance in a given pixel in $\rm MJy/sr$, $\rm e_{\rm gain}$ is the gain between electrons and digital numbers in $\rm e^- / DN$ and $t$ is the exposure time in seconds. Using these expected electron counts in each pixel we perform a single Poisson sample for each pixel to determine its electron count, then convert back to $\rm MJy/sr$. We used a value of 2.0 for $\rm e_{\rm gain}$ for all filters. The $\rm sens_{\rm avg}$ value for each filter is determined by averaging the "PHOTMJSR" values for each filter found in the JWST calibration and zeropoint table as of October 2022 \citep{Gordon_2022}\footnote{Table is available at \url{https://jwst-docs.stsci.edu/files/182256933/182256934/1/1669487685625/NRC_ZPs_0995pmap.txt}.}. The resulting $\rm sens_{\rm avg}$ values for the F444W, F356W, F277W, F200W, and F150W filters are $0.383$, $0.391$, $0.453$, $1.914$, and $2.255$ $\rm (MJy/sr) / (DN/s)$ respectively. The distinct jump between the F277W and F200W filters is due to the smaller pixel size of the short wavelength filters compared to the long wavelength filters.}

Finally, we drizzle the three images of each galaxy together using the \texttt{drizzle} python package, which is based on \texttt{DrizzlePac}. We use the same drizzle parameters as CEERS which results in images with pixel sizes of 0.030" for both the long and short wavelength filters.\review{After performing this drizzle, we subtract the expected background radiance of $0.25 \rm MJy/sr$ from the image to offer a background subtracted image in addition to the raw observation. The background subtracted images are used for all following analysis in this work.}

This process results in mock observations like the individual galaxies shown in fig. \ref{fig:galaxy_grid} and the galaxy cluster fig. \ref{fig:Group_image}. The false color images presented in those figures are made by assigning 3 different NIRCam filters to each of the red, green, and blue channels of the image. We present the F356W, F277W, and F150W as the red, green, and blue channels respectively. Figure \ref{fig:image_decomp} shows the false color images of two galaxies, along with black and white renditions of each of the filters that are present. It is apparent that the F356W and F277W observations are much smoother than the F150W observations. This is primarily due to the fact that both the F277W and F356W filters are long-wavelength filters, so their pre-drizzle resolution is 0.063" and their PSF is twice that of the short-wavelength filters, and thus get smoothed significantly when being drizzled to 0.030", while the F150W filter is a short wave-length filter that has a pre-drizzle resolution of 0.031", and a significantly tighter PSF, so minimal smoothing occurs during the drizzling process.

\section{Results}
\label{sec:results}
We use the above imaging pipeline to create synthetic images of $\sim 350,000$ subhalos from redshift 3 to 6 with five NIRCAM filters (F150W, F200W, F277W, F356W, F444W). We measure a number of morphological and size properties of these galaxies in order to compare this dataset to existing observations, and similar simulation based data sets, and discern as much as possible about the morphologies of galaxies in this redshift range.

\begin{figure*}
    \centering
    \includegraphics[width=1.0\columnwidth]{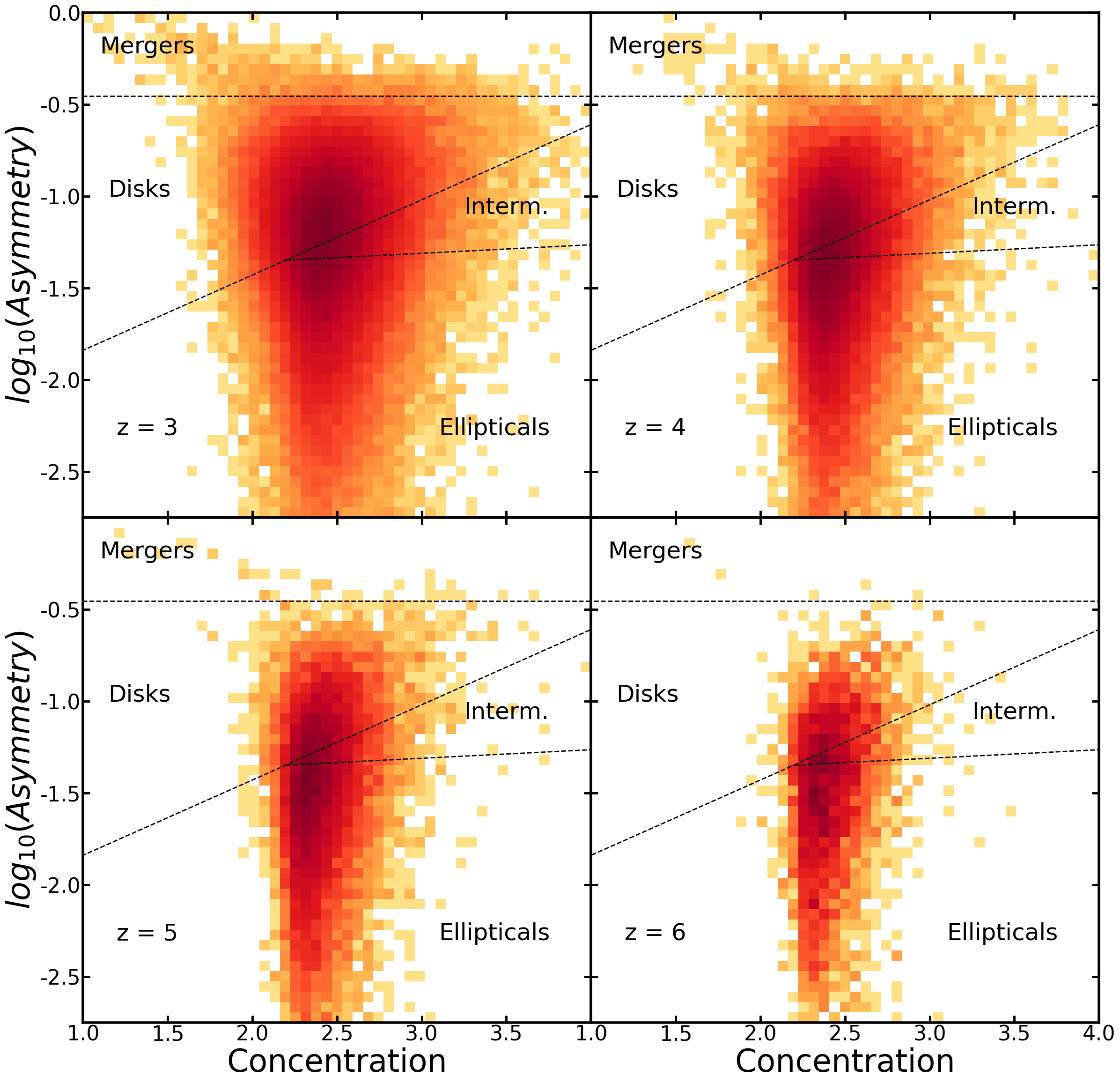}
    \includegraphics[width=1.0\columnwidth]{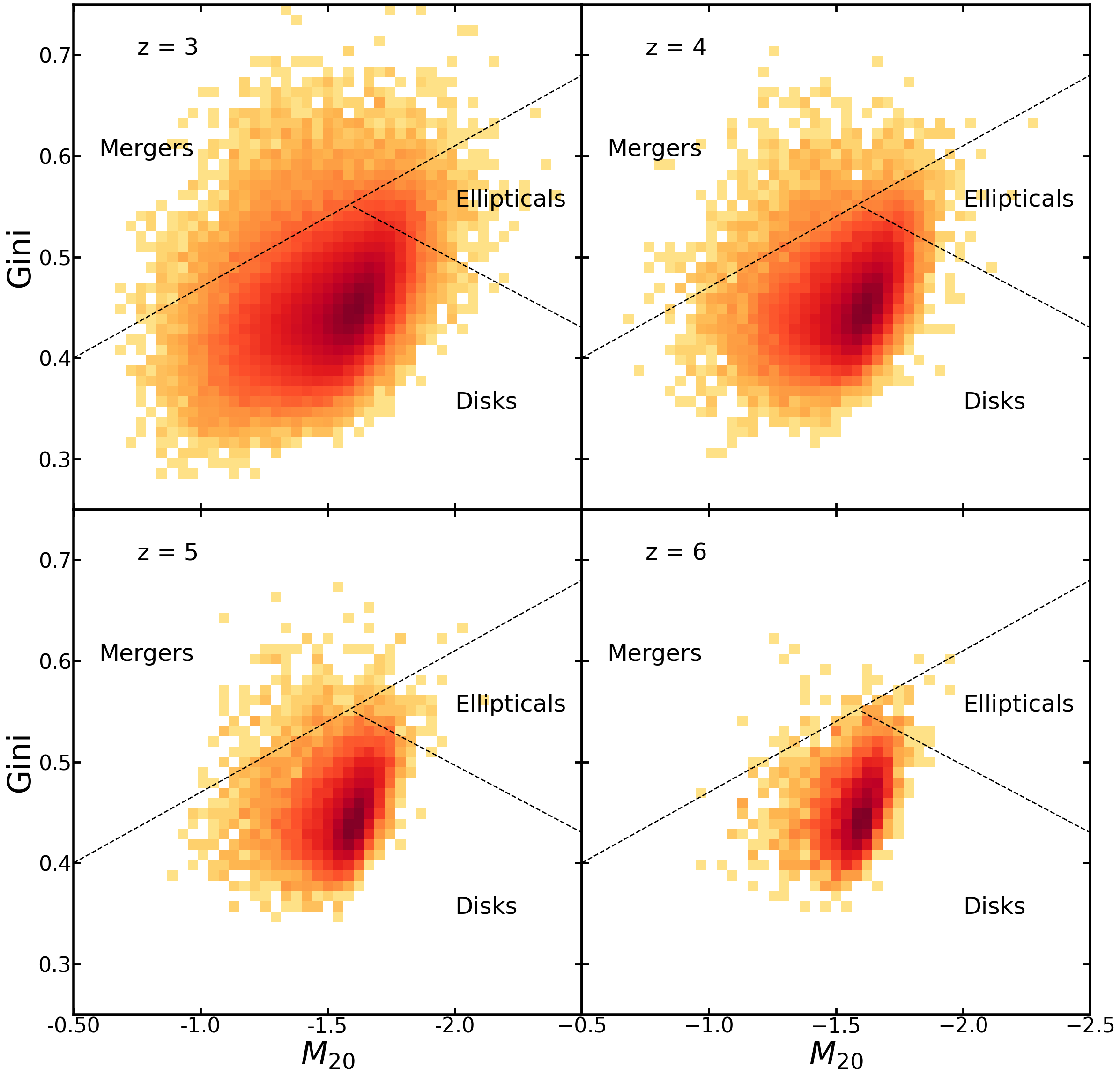}
    \caption{The left panel shows the Asymmetry-Concentration density plots for each redshift. The dashed lines indicate the the boundaries between elliptical, intermediate, disk, and merger galaxies detailed in \citet{Bershady_00} and \citet{Conselice_03}. The right panel shows Gini-M20 density plots for each redshift. The dashed lines indicate the boundaries between disk galaxies, elliptical galaxies, and galaxies undergoing mergers shown in \citet{Lotz_08}.}
    \label{fig:CAS_Gini_M20}
\end{figure*}

\subsection{Morphological Measures}
\label{sec:morph}

We use the \texttt{statmorph} code package \citep{statmorph} to measure the parametric and non-parametric morphological properties of each galaxy. \review{We use this package in its standard configuration as laid out in \citet{statmorph}, and exclude all galaxies that had a fit or Sersic fit flag of any kind, and a signal-to-noise ratio below 2.5 from our analysis.} In this work we largely focus on the parameters of the Sersic fit, and the Concentration, Asymmetry, and Smoothness (CAS) statistics \citep{Bershady_00, Conselice_03}, as well as the Gini coefficient and $M_{20}$ value \citep{Abraham_03, Lotz_04}.

\subsubsection{Sersic Fitting}
\label{sec:Sersic_fitting}

\begin{figure*}
    \centering
    \includegraphics[width=2.0\columnwidth]{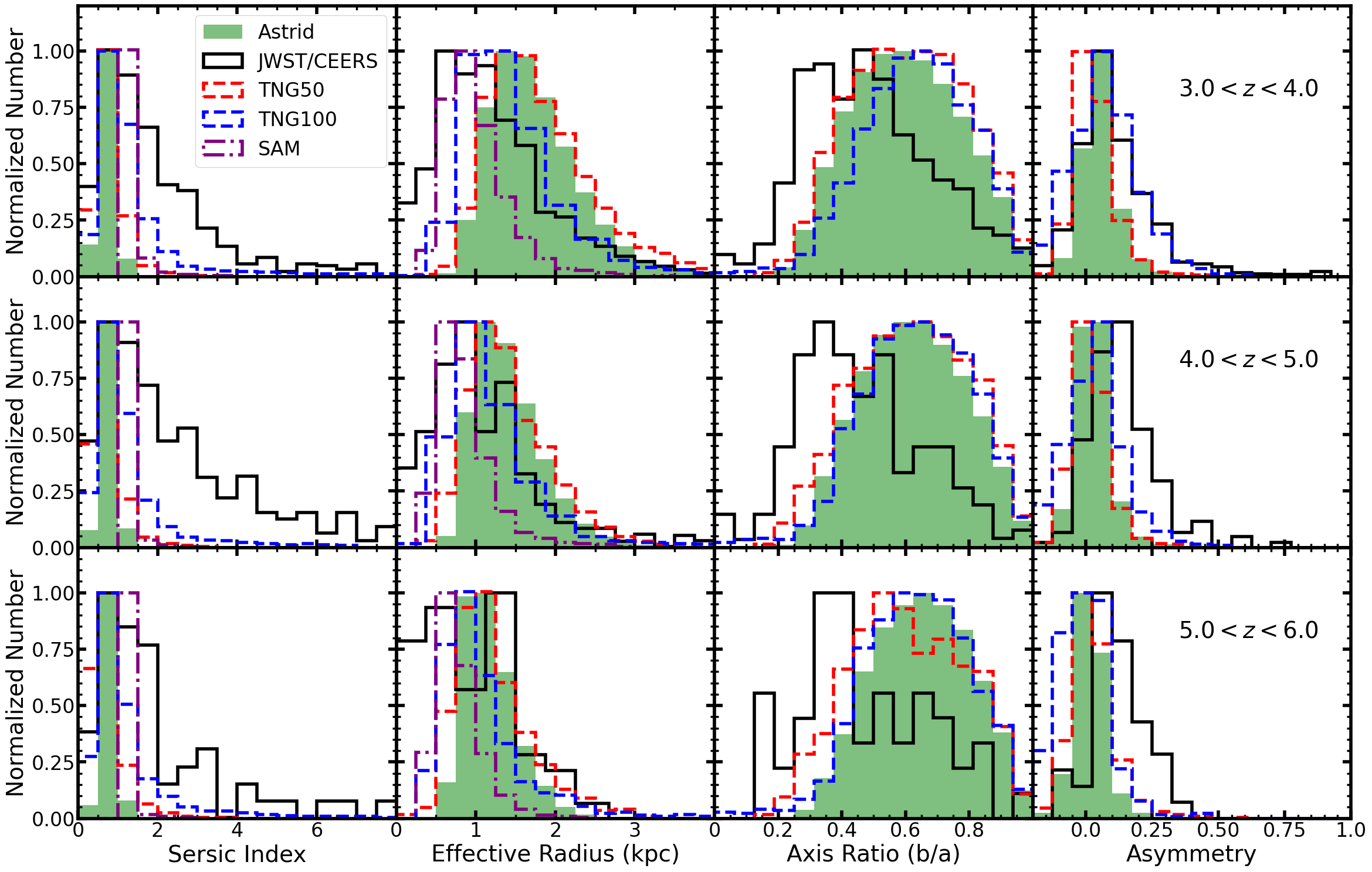}
    \caption{Distributions of Sersic index, Sersic effecive radius, Axis Ratio, and Asymmetry for galaxies in the Astrid simulation, compared with the distributions for CEERS data release one \citep{CEERS_3}, mock images from Illustris TNG50 \citep{TNG50_CEERS} and TNG100 \citep{Rose_23}, and mock CEERS images produced using a SAM \citep{Somerville_2015, Somerville_2021, Yung_2019, Yung_2022}. In order to approximate the redshift ranges indicated in the figure for the Astrid catalog, which only contaizns integer redshifts, we used the galaxies from the bounding redshifts.}
    \label{fig:CEERS_TNG_comp}
\end{figure*}

The distribution of Sersic indices for our galaxy sample is shown in Figure \ref{fig:Sersic_hist}. \review{We can see that the overall distribution remains largely unchanged across the redshift range in our sample}. Traditionally, Sersic indices near  $n=1$ are associated with disk galaxies, and larger Sersic indices are associated with elliptical galaxies ($n=4$ produces the de Vaucouleurs profile). \review{The distribution of Sersic indices for our entire sample has a mean of $n=0.72$ and $\sigma = 0.21$, with only $0.10\%$ of galaxies having a Sersic index of 2 or higher.} Observational datasets for galaxies at these redshifts indicate that disk galaxies should have a narrow distribution centered near $n=1.0$ while elliptical galaxies should have a broader distribution centered between $n=2.0$ and $n=3.0$ \citep{CEERS_3, Vika_2015, Kartaltepe_2015}. This indicates we may be capturing elliptical galaxies that fall on the lower end of the broad distribution, but are likely missing many of the higher Sersic index galaxies. That said, our sample matches quite well with other samples made from simulations. \review{In the first column of figure \ref{fig:CEERS_TNG_comp} we can see our distribution of Sersic indices compared with other samples including those made form Illustris TNG50 and TNG100 \citep{TNG50_CEERS, Rose_23}, and our distribution is comparable to theirs, while being slightly narrower}.

There is also valuable galaxy size information that can be gleaned from the Sersic fit via the effective radius. In Figure \ref{fig:Sersic_hist} we show the distribution of Sersic effective radii for galaxies in our sample, and summary statistics for those distributions can be found in table \ref{table:sizes}. There is a marked increase in the mean effective radius between each redshift from $z=6$ to 3, which is expected. A number of studies at redshifts below $z=3$ have shown the same relation between redshift and galaxy size \citep{ribeiro_16, trujillo_06, vanderwel_14, Shibuya_2015} and extrapolating those relations to the redshifts in our sample shows the same relation. There are also initial results from JWST observations  (\citealt{Ormerod_2023}) that show the same size evolution within our redshift range.

\review{
\begin{table}
\caption{Distribution statistics for the effective radii of galaxies at each redshift in our sample. 16th and 84th percentiles are shown as modifiers to the median values.}
\centering
{\renewcommand{\arraystretch}{1.5}
\begin{tabular}{c|c|c|c}
z & mean $r_e$ (kpc) & $\sigma_{r_e}$ (kpc) & median $r_e$ (kpc) \\
\hline
3 & 1.85           & 0.63     & $1.75^{+0.65}_{-0.47}$                 \\
4 & 1.50           & 0.47     & $1.42^{+0.51}_{-0.36}$                 \\
5 & 1.22           & 0.35     & $1.15^{+0.39}_{-0.27}$                 \\
6 & 1.06           & 0.29     & $1.01^{+0.33}_{-0.22}$                
\end{tabular}
}
\label{table:sizes}
\end{table}
}

\subsubsection{Non-Parameteric Measures
\label{sec:non-parametric_measures}}

In addition to the Sersic fit, we measure some nonparametric morphology measures. First, we examine the measurements of the CAS statistics for our galaxy sample. In the left panel of Figure \ref{fig:CAS_Gini_M20} we show the distribution of our galaxy sample (separated by redshift) on the Asymmetry-Concentration plane, including the traditional boundaries between elliptical, intermediate, and disc galaxies, along with galaxies undergoing mergers \citep{Bershady_00, Conselice_03}. Based on those boundaries our sample should include significant amounts of elliptical, intermediate, and disk galaxies, and very few mergers. This deficit of galaxies classified as mergers based on these boundaries is also seen in \citet{CEERS_3}.

We also measure the Gini coefficient, and the second moment of the central region containing $20\%$ of light ($M_{20}$). In the right panel of Figure \ref{fig:CAS_Gini_M20} we show our galaxy population on the Gini-$M_{20}$ plane (separated by redshift), including the traditional boundaries between disk galaxies, elliptical galaxies, and mergers. Here we see a strong bias toward the disk portion of the plot, similar to our Sersic index results. That said, there is a good deal of work \citep{Pearson_2019, Kartaltepe_2010} including \citet{CEERS_3} that indicates populations of disk and elliptical galaxies in this redshift range show very little difference on these axes. \review{The boundaries for both the Gini-$M_{20}$ plane, and the Asymmetry-Concentration plane were determined based on observations of low redshift galaxies, so they may need to be shifted substantially in order to properly divide galaxies at higher redshifts, as indicated by their poor seperation of observations noted in \citet{Pearson_2019, Kartaltepe_2010, CEERS_3}. With the increase in high fidelity galaxy images at high redshift from JWST, it may be possible to determine what those shifts should be.}

\subsubsection{Comparison with other datasets}
\label{sec:dataset_comparisons}

We compare the results of our morphological fits to those of CEERS \citep{CEERS_3}, the CEERS mock catalogs produced from Illustris TNG \citep{TNG50_CEERS,Rose_23}, and the mock catalog produced by the Santa Cruz Semi-analytical Model \citep{Somerville_2015, Somerville_2021, Yung_2019, Yung_2022} based on their presentation in \citet{CEERS_3} in Figure \ref{fig:CEERS_TNG_comp}. Overall the \texttt{Astrid} galaxy sample has similar results to the samples made using Illustris TNG50 and TNG100, and shares many characteristics with the CEERS results, but with some key deviations. \review{The distribution of Sersic indices for \texttt{Astrid} is similar, but more concentrated than the TNG50 and TNG100 distributions, all of which peak at lower values, and are far narrower, than the CEERS observational data.} Similarly, the Astrid, and TNG size distributions are comparable, but are biased towards larger galaxies than the CEERS dataset. These discrepancies in both Sersic index, and effective radius are likely linked indicating that the simulated galaxies have flatter, more diffuse light profiles than the observed galaxies. We examine this effect further in section \ref{sec:dust_and_instrumentation}. 

In the third column of Figure \ref{fig:CEERS_TNG_comp} we compare the axis ratios of these datasets. \review{While the datasets match well overall, the CEERS dataset peaks at a lower value than the simulation datasets which are all comparable. In the fourth column we see that the \texttt{Astrid} dataset is still  comparable to the TNG datasets, all of which have a larger proportion of galaxies with negative asymmetry values than the CEERS dataset.} A negative asymmetry value indicates very little about the galaxy itself, but is rather the effect of a low signal-to-noise ratio on the measurement of the source. The shared bias towards lower asymmetry among the simulation-based samples indicates a potential deficit in the ability of simulations to produce galaxies with features that result in high asymmetry.

When comparing to these datasets we check the impact that the galaxy stellar mass function (GSMF) and redshift distribution has on the result, as our population has a slightly different GSMF than the CEERS dataset itself, as seen in figure \ref{fig:gal_sample}, and our population is confined to integer redshifts rather than having a continuous redshift distribution. We weight our dataset by galaxy mass and redshift in order to match the CEERS sample as closely as possible in those two metrics and compare with our results shown in figure \ref{fig:CEERS_TNG_comp}. \review{We see minimal shifts in all metrics. For example the mean Sersic index in the redshift 3 to 4 band shifts by 0.01 after weighting.}

\begin{figure*}
    \centering
    \includegraphics[width=2.0\columnwidth]{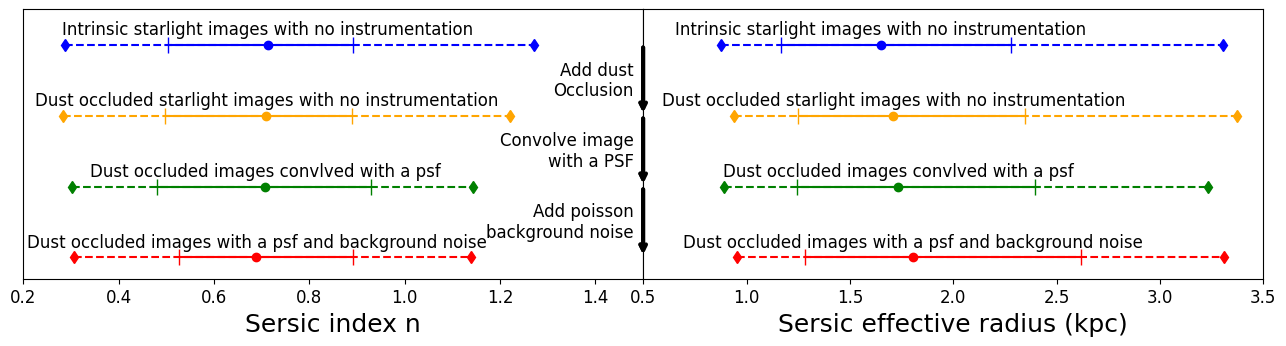}
    \caption{The distribution of the Sersic index and effective radius at different stages of the mock observation process. Central points are the median value for that step, the solid lines and end caps show the 16th and 84th percentiles of the distribution, while the dashed lines and diamonds show the $2\sigma$ (2.5 and 97.5 percentiles) points of each distribution.}
    \label{fig:image_stages}
\end{figure*}

\subsection{The impact of mock observation steps on the morphological measures}
\label{sec:image_stages}

In order to examine the impacts of each element in our imaging pipeline we create mock observations of a subset of our galaxy sample at a few key points in the pipeline and fit those observations with Statmorph to see how the distributions of some of our morphological measures are impacted by each element. We choose $\sim 4000$ galaxies in the redshift $z=3$ snapshot to examine, and use the F277W filter for analysis. We examine two specific portions of the pipeline in-depth. First, we examine the impact of adding dust occlusion 
and how elements of the instrument simulation affect  the distribution of Sersic indices and effective radii which can be found in figure \ref{fig:image_stages}. Second, we compare fits from single $t=945 s$ exposures, and observations made from drizzling 3 such exposures together to analyze the impact that increased exposure time, and drizzling have on our sample.

\subsubsection{Dust Occlusion and Instrumentation}
\label{sec:dust_and_instrumentation}

For our analysis of dust occlusion and instrumentation we choose the following imaging configurations: Intrinsic starlight with no dust occlusion, and no instrumentation (PSF or noise) as a baseline before any of our dust modelling, or instrumentation have been added. Next we make images after the dust occlusion is added onto those images but still without instrumentation to examine the impact of the dust model exclusively. Then we examine those images after they are convolved with a PSF, and finally the galaxy images after background poisson noise has been added. We perform multiple exposure drizzling for all of these stages for consistency.

The primary conclusion is that these steps each have a minimal impact overall on the distribution of the Sersic index and radius for these galaxies. More prominent effects could likely be seen in statistics like the asymmetry, where high background noise is the driving factor for the asymmetry values below 0 discussed earlier in section \ref{sec:dataset_comparisons}.

That said, there are important trends that can be gleaned from this analysis. We see the addition of dust occlusion slightly lowers the Sersic index of our sample, and increases the effective radius. This is expected, as the dust occlusion may dim the very bright and dense centers of galaxies, flattening their light curves, and increasing the relative brightness of the edges of galaxies, which results in a lower  Sersic index, and a larger effective radius. \review{This effect is not evident in the median of the distribution of Sersic indices as it is equal for both stages, but rather in the position of the $2\sigma$ point which decreases from $n=1.27$ to $n=1.22$ between these two stages.} This indicates that the flattening effect of dust occlusion is most pronounced on the galaxies that are more elliptical.

The convolution of the PSF results in a slight decrease in the median and overall distribution of the effective radius, and a slight increase in the median Sersic index, along with an overall narrowing of the distribution. These effects are the opposite of what one would be expect for the addition of a PSF, given that it generally smooths out images. However,  the PSF is also passed to \texttt{Statmorph} so it can be accounted for during the fitting process. This process of adding the PSF, and then attempting to deconvolve it during fitting results in slightly smaller, more peaked images.

The addition of noise results in another slight decrease in the median of the distribution of the Sersic index, and a significant increase in the effective radius. This is likely due to the noise blurring the edges of a galaxy, making it appear to taper off slower, which corresponds to a light profile with a lower Sersic index, and a larger effective radius.

\begin{figure*}
    \centering
    \includegraphics[width=1.0\columnwidth]{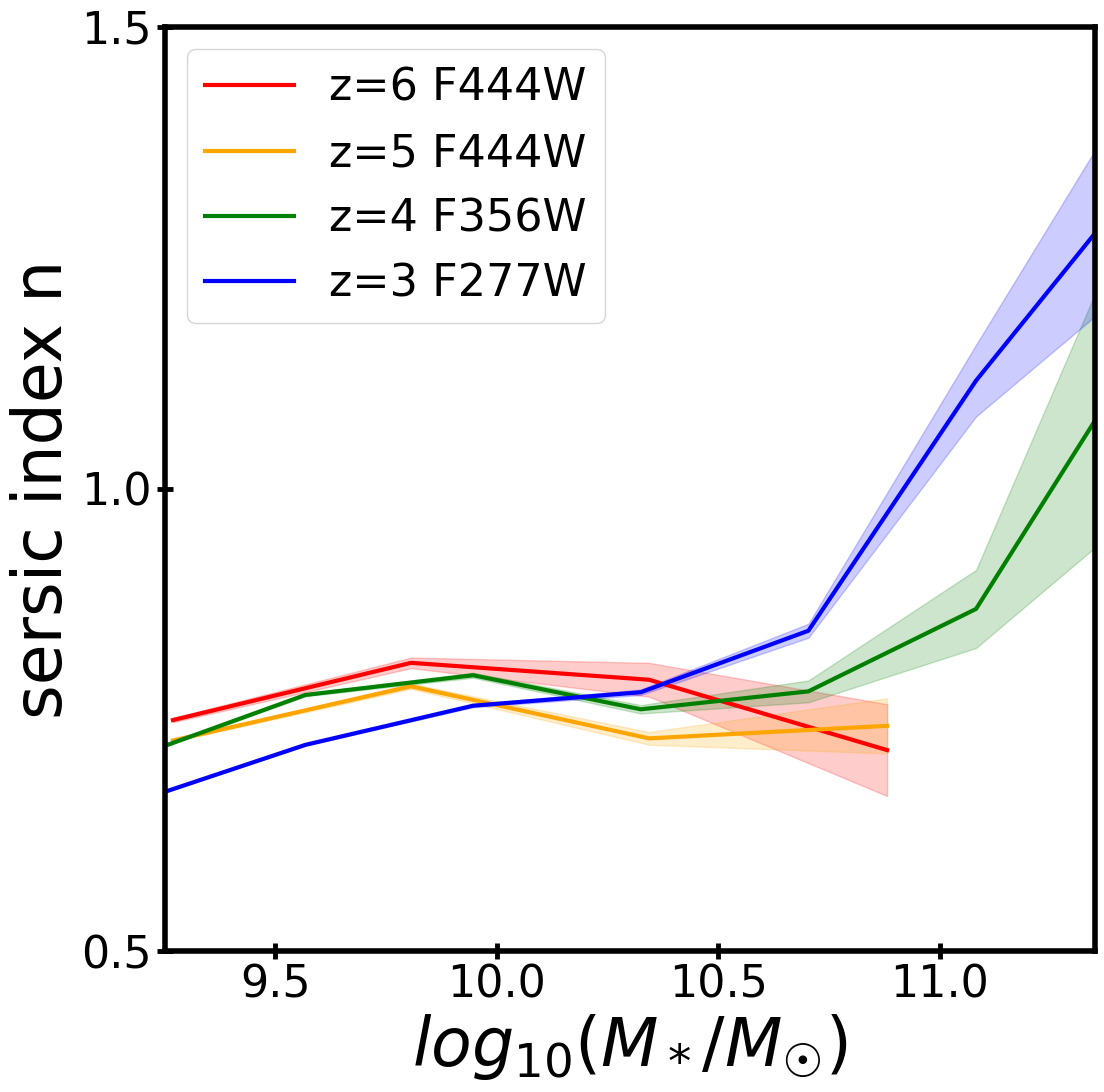}
    \includegraphics[width=1.0\columnwidth]{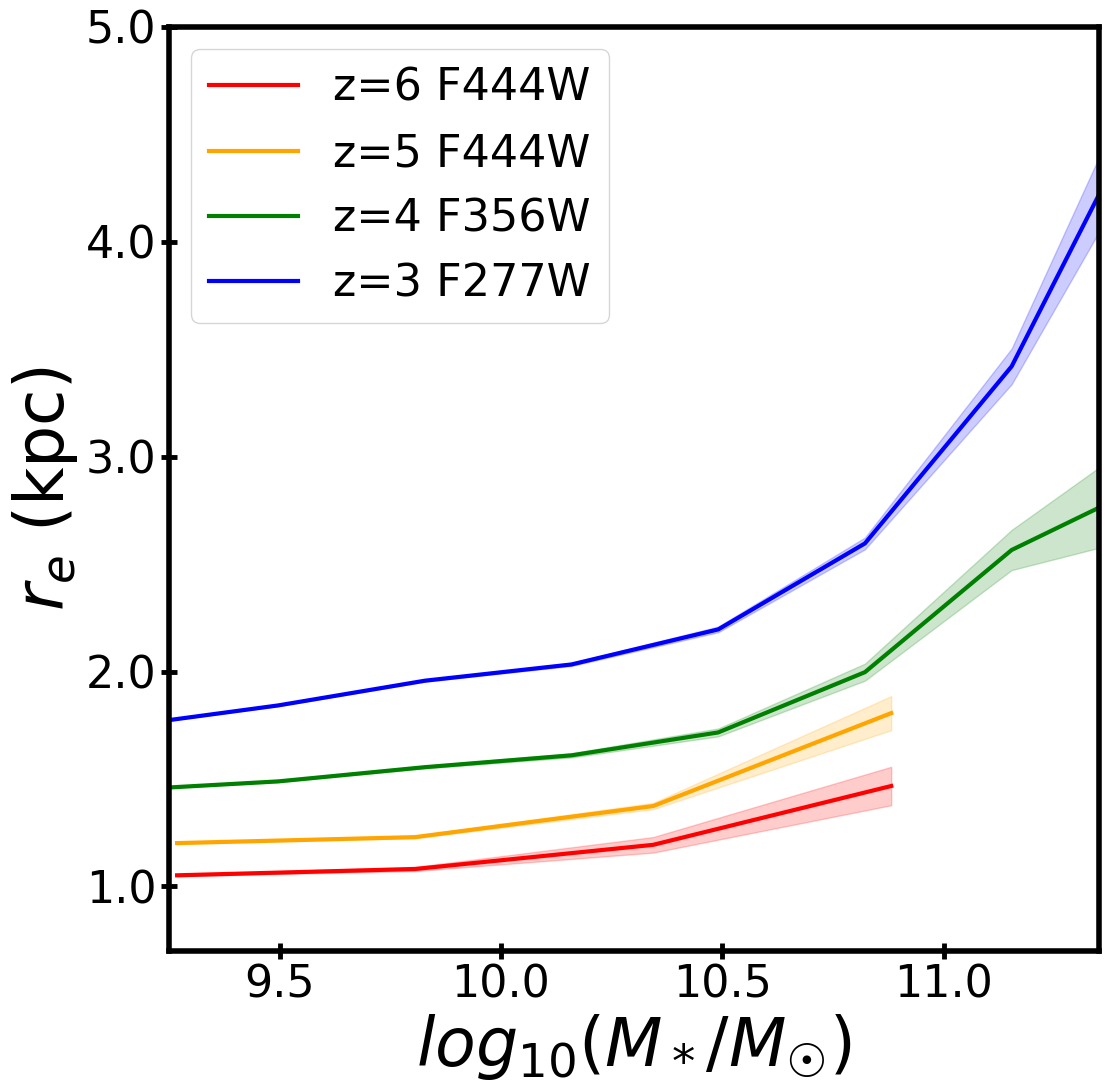}
    \caption{Left Panel: Sersic Index-Galaxy Mass relations for redshifts $z=3-6$. Right Panel: Stellar Mass-Sersic Effective Radius relations for redshifts $z=3-6$. Shaded regions around each line indicate the uncertainty on the mean in each bin.}
    \label{fig:Sersic_vs_SM}
\end{figure*}

\subsubsection{Observation Drizzling}
\label{sec:drizzling}

We also examine the impact that drizzling and exposure time have on these distributions by fitting single exposures with exposure time $t=945 s$ compared to the drizzled images made from 3 exposures. \review{For the subsample we examine this has a negligible effect on the mean of the distribution for both Sersic index ($n= 0.74 \pm 0.27, n=0.71 \pm 0.22$ for undrizzled and drizzled images respectively) and the Sersic effective radius ($r_e = 1.95 \pm 0.79 \rm kpc, r_e = 1.95 \pm 0.74 \rm kpc$ respectively).} On the other hand, the undrizzled images are slightly more likely to have issues during the fitting process. \review{99\% of the drizzled observations are able to be fit without issue, while only 96\% of the undrizzled observations of the same galaxies are properly fit. All of the galaxies that are well fit using the drizzled observations, but are unable to be fit using the undrizzled observations have a mass of $\rm log{M_*/\msun} = 9.7$ or less.} This shows the value of the increased sensitivity of JWST’s instruments, and longer exposure drizzled observations for capturing lower mass dim galaxies.

The minimal impact of each of these imaging steps, and the stark differences between the distributions of morphological parameters from observational samples and samples created from cosmological simulations indicate that the majority of the difference between the observations and mock catalogs is likely a result of limitations of the current generation cosmological simulations rather than the choices made in the mock observation pipelines. The most obvious discrepancy can be seen from the Sersic fits alone. Our dataset, and both datasets created using the Illustris TNG simulations lack the high Sersic index, and low radius galaxies seen in observations like CEERS. There are a number of properties of current cosmological simulations that could be contributing to this, but the most likely is that the high mass resolution, and limited spatial resolution of large cosmological simulations limits their ability to create the very dense regions necessary to produce galaxies that have higher Sersic indices and lower effective radii.

\section{Analysis}
\label{sec:analysis}

In this Section we leverage the large number of galaxies in our sample in order to analyze how some of the morphological measures vary with mass and redshift. Additionally, we compare the galaxy classifications from Sersic fitting, the CAS statistics, and the Gini-$M_{20}$ measures. Finally, we examine the potential impact of adding the light produced by Active Galactic Nuclei (AGNs) on our mock images.

\begin{table}
\caption{Coefficients and errors for the linear regression of the mass-Sersic index relation of different redshifts shown in the left panel of figure \ref{fig:Sersic_vs_SM}.}
\centering
\centering
{\renewcommand{\arraystretch}{1.5}
\begin{tabular}{c|cccc|cccc}
z & \multicolumn{4}{c|}{All Masses}                            & \multicolumn{4}{c}{$log_{10}(M_*/\msun) \geq 10.25$}       \\
\hline
  & $\alpha_n$ & $\beta_n$ & $\sigma_{\alpha}$ & $\sigma_{\beta}$ & $\alpha_n$ & $\beta_n$ & $\sigma_{\alpha}$ & $\sigma_{\beta}$ \\
3 & 0.142   & 0.638  & .001             & 0.001           & 0.268    & 0.452   & 0.010             & 0.013            \\
4 & 0.097   & 0.701   & 0.002           & 0.001           & 0.058    & 0.703   & 0.014             & 0.019            \\
5 & 0.079   & 0.710   & 0.004          & 0.002            & -0.047   & 0.804   & 0.028             & 0.036            \\
6 & 0.112   & 0.722   & 0.008          & 0.003            & -0.181   & 1.003   & 0.072             & 0.089                 
\end{tabular}
}
\label{table:n_vs_SM}
\end{table}

\begin{table}
\caption{Coefficients and errors for the linear regression of the mass-size relation of different redshifts show in the right panel of figure \ref{fig:Sersic_vs_SM}.}
\centering
{\renewcommand{\arraystretch}{1.5}
\begin{tabular}{c|cccc|cccc}
z & \multicolumn{4}{c|}{$log_{10}(M_*/\msun)$}                            & \multicolumn{4}{c}{$log_{10}(M_*/\msun) \geq 10.50$}       \\
\hline
  & $\alpha_r$ & $\beta_r$ & $\sigma_{\alpha}$ & $\sigma_{\beta}$ & $\alpha_r$ & $\beta_r$ & $\sigma_{\alpha}$ & $\sigma_{\beta}$ \\
3 & 0.378    & 1.678   & 0.004             & 0.002            & 1.648    & -0.18   & 0.045             & 0.07             \\
4 & 0.196    & 1.413   & 0.005             & 0.003            & 0.956    & 0.36    & 0.062             & 0.09             \\
5 & 0.119    & 1.174   & 0.007             & 0.003            & 0.846    & 0.28    & 0.11              & 0.17             \\
6 & 0.103    & 1.027   & 0.013             & 0.006            & 0.882    & -0.02   & 0.25              & 0.37                 
\end{tabular}
}
\label{table:reff_vs_SM}
\end{table}

\subsection{The evolution of morphological measures with redshift}
\label{sec:morph_analysis}

We focus on examining the Sersic index and effective radius when analyzing the redshift evolution of our morphological fits. Specifically, we analyze how their relationships with stellar mass evolve with redshift, as that can provide insights into galaxy evolution.

\subsubsection{Morphology-Mass and Size-Mass Relations}
\label{sec:mass_relations}

In the left panel of figure \ref{fig:Sersic_vs_SM} we show the Sersic index as a function of galaxy mass for each redshift. We perform a linear regression on the population at each redshift to find the relationship between $n$ and mass using the following

\begin{equation}
n = \alpha_n *(log_{10}(M_*/\msun) - 9) + \beta_n,
\end{equation}

and the results can be found in table \ref{table:n_vs_SM}. There is a significant change in the relationship between $n$ and mass as time progresses. \review{The relationship is consistently positive, and generally increases as redshift decreases ($\alpha_n = 0.112, 0.079, 0.097,$ and $0.142$ for $z=6, 5, 4, $ and $3$ respectively). These relations are the most drastic at the high mass end, as the relations are largely flat for galaxies with $log(M_*/\msun) \leq 10.0$, and the high number of galaxies at the low mass end tends to dominate the fit.} The high mass behavior exhibits a consistent trend of galaxies in the same mass bin having higher Sersic indices at lower redshift, indicating that the morphology of high mass halos evolves over this redshift range. We perform the same linear fit outlined above on the population of galaxies with $log(M_*/\msun) \geq 10.25$ to quantify this evolution. This can also be seen in table \ref{table:n_vs_SM}. Specifically this indicates that the population of high mass galaxies contains more disks and irregular galaxies at high redshift, and that the population shifts toward more elliptical galaxies as time progresses.

\review{There is a great deal of work using observations \citep{Whitney_2021, Bluck_2012, Tamburri_2014}, simulations \citep{Rodriguez-Gomez_2015}, and semi-empirical models \citep{Hopkins_2010} that show galaxy merger rates increase with redshift  which would lead to more high mass irregular galaxies with low Sersic indices at higher redshifts.} There is also work that shows this same trend of high mass galaxies having lower Sersic indices and being more disk-like at higher redshifts \citep{Buitrago_2013, Ormerod_2023, Shibuya_2015}. Both of these effects would lead to lower average Sersic indices for high mass galaxies at higher redshifts, so the presence of that trend in our data supports them but this analysis alone is not enough to differentiate between the two.

In the right panel of figure \ref{fig:Sersic_vs_SM} we show the effective radius as a function of galaxy mass for each redshift. We perform the same linear regressions on this dataset as we did on the Sersic index data set using the following:
\begin{equation}
r_e = \alpha_r *(\log_{10}(M_*/\msun) - 9) + \beta_r,
\end{equation}
with the results shown in Table \ref{table:reff_vs_SM}. We see a positive correlation between mass and effective radius for all four redshifts in our sample, with the correlation coefficient increasing as redshift decreases. This positive correlation between mass and radius is in agreement with observational studies of galaxy size in our redshift range \citep{Bouwens_2022, Shibuya_2015, Mosleh_2020, Roy_2018}. That said, some simulation studies   \citep{Marshall_2022, Roper_2022} and some theoretical predictions \citep{Tacchella_2016, Tacchella_2015, Costantin_2021} point to the formation of very compact massive galaxies in this redshift range that could result in a negative relationship between galaxy mass and size. While no redshift in our sample shows this negative relationship, the correlation does increase from redshift 6 to 3. This may indicate the mechanism that would result in a negative mass-size relation would be more dominant at higher redshifts, while the mechanisms that result in a positive mass-size relation have increasing impact as redshift decreases.

Another noticeable feature of the $r_e$ vs mass plot is that all 4 redshifts have a fairly flat relationship between mass and radius for galaxies below $\log(M_*/\msun) = 10.5$, and display an upturn after that point. This upturn is  noticeable for $z=4-6$ but becomes most pronounced for $z=4$. Similar upturns at the high mass end are seen in observations and supported by theory regarding the growth mechanisms for galaxies at different masses \citep{Nedkova_2021, Mosleh_2020, Shen_2003, Mosleh_2013, Mowla_2019}.

\subsubsection{Comparison of light-based and mass-based size measures}
\label{sec:sizes}

\begin{figure}
    \centering
    \includegraphics[width=1.0\columnwidth]{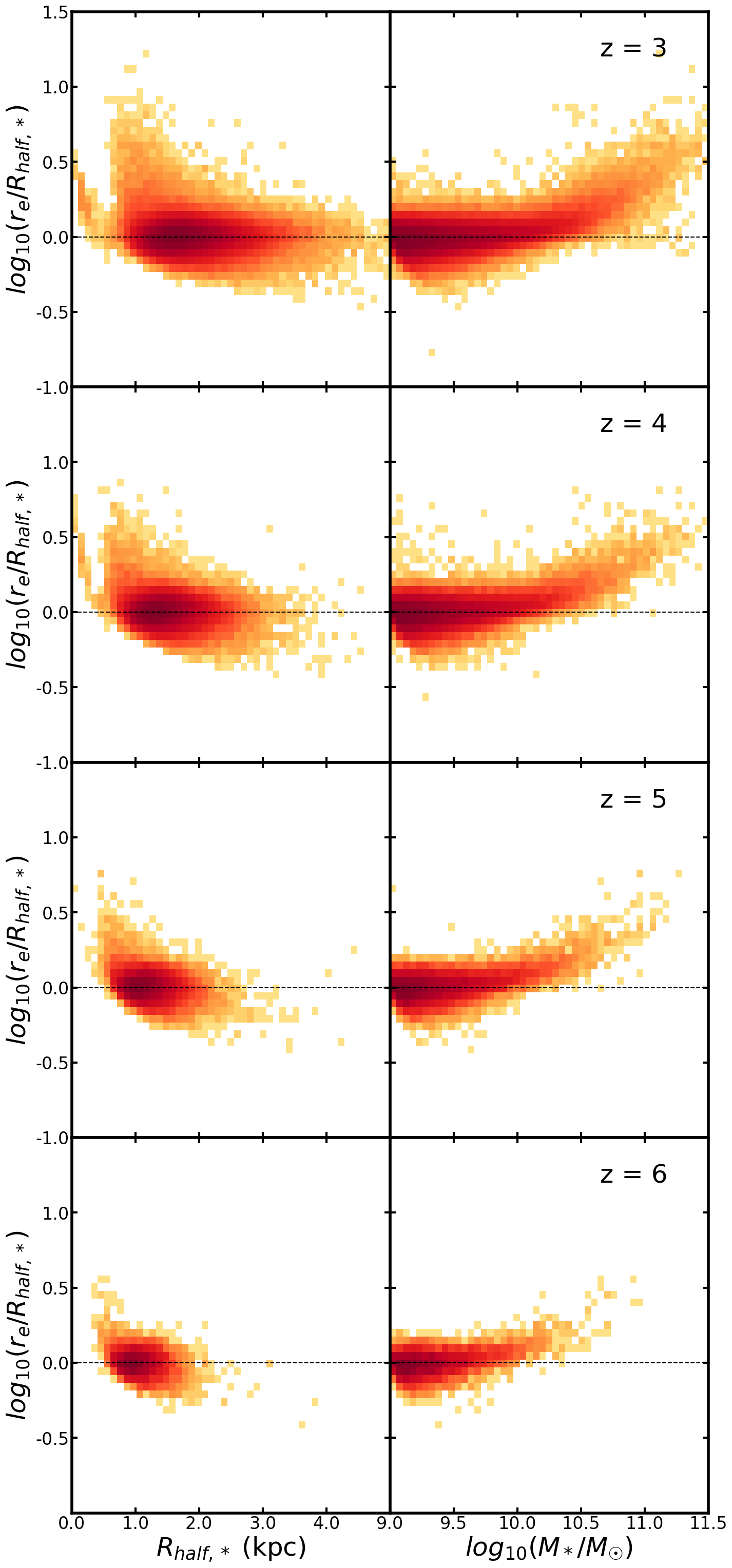}
    \caption{A comparison between the stellar half mass radius $R_{\rm half,*}$ measured from the simulation, and the Sersic effective radius $r_e$ from the Sersic fit. The left column shows the ratio of the two radii as a function of $R_{\rm half,*}$ indicating how accurate $r_e$ is as a proxy for the half mass radius for different sized galaxies. The right column shows the radius ratio as a function of galaxy mass to identify if galaxy mass is correlated with deviations from the $r_e \sim R_{\rm half,*}$ relation. Each row shows these relations for different redshift snapshots from z=3 in the top row to z=6 in the bottom row}
    \label{fig:reff_vs_rhalf}
\end{figure}

Additionally, we compare the effective radius to the stellar half-mass radius in the left half of Figure \ref{fig:reff_vs_rhalf} to evaluate how useful the Sersic effective radius is as a proxy for the half-mass radius as was done by \citet{TNG50_CEERS}. We report the median $R_{\rm half,*}$, $r_e$, and their ratio in table \ref{table:radius_ratios}. Similar to \citet{TNG50_CEERS} we find that $r_e/R_{\rm half,*} \approx 1$ at all redshifts. There is a noticeable outlier region present at $R_{half_*} \approx 1 kpc$ for all redshifts, where many galaxies show a ratio well above 1. This indicates there is a population of galaxies with fairly small half mass radii that have systematically too large effective radii.

In order to investigate this we plotted $r_e/R_{\rm half,*}$ vs galaxy stellar mass for each redshift in the right half of the figure. While there are galaxies of all masses that deviate from the 1-to-1 relation, there is a consistent trend of higher mass galaxies having coefficients well above 1. This could be due to the Sersic fit systematically overestimating some class of high mass galaxies, or some portion of our pipeline not properly representing the light from galaxies.

Our pipeline has elements that can lead to the over estimation of the effective radius, and the underestimation of the Sersic index as discussed in section \ref{sec:dust_and_instrumentation} and earlier in this section. Given these are massive galaxies that are quite bright it is unlikely that this is the result of edge blurring due to background noise, but it is possible our dust model is over occluding light from the centers of these galaxies resulting in flatter light profiles that get fit with larger effective radii. There is some evidence for this in Figure \ref{fig:image_stages} as the dust-occluded effective radii are larger than those from observations without dust. While the effect on the overall distribution is small individual high mass galaxies might see a significantly larger impact from this effect. Despite that, a similar trend is present when the same analysis is performed in \citet{TNG50_CEERS} where an entirely different dust model and imaging pipeline are used. This indicates there is likely something more fundamental within the simulations that are causing this effect, or some element of galaxy morphology that results in a mismatch of the half light and half mass radii.

\begin{table}
\caption{Summary statistics for intrinsic and observed sizes, along with their ratio. 84th and 16th percentile values are presented via their difference from the median}
\centering
{\renewcommand{\arraystretch}{1.5}
\begin{tabular}{c|ccc}
z & $R_{\rm half,*}$ (kpc)    & $r_e$ (kpc)           & $r_e / R_{\rm half,*}$ \\
\hline
3 & $1.74^{+0.54}_{-0.39}$ & $1.75^{+0.65}_{-0.47}$ & $1.00^{+0.18}_{-0.15}$       \\
4 & $1.41^{+0.41}_{-0.31}$ & $1.42^{+0.51}_{-0.36}$ & $1.01^{+0.18}_{-0.15}$        \\
5 & $1.15^{+0.32}_{-0.25}$ & $1.15^{+0.39}_{-0.27}$ & $1.02^{+0.17}_{-0.15}$       \\
6 & $1.02^{+0.28}_{-0.21}$ & $1.01^{+0.33}_{-0.22}$ & $1.00^{+0.17}_{-0.14}$                        
\end{tabular}
}
\label{table:radius_ratios}
\end{table}

\begin{figure*}
    \centering
    \includegraphics[width=\textwidth]{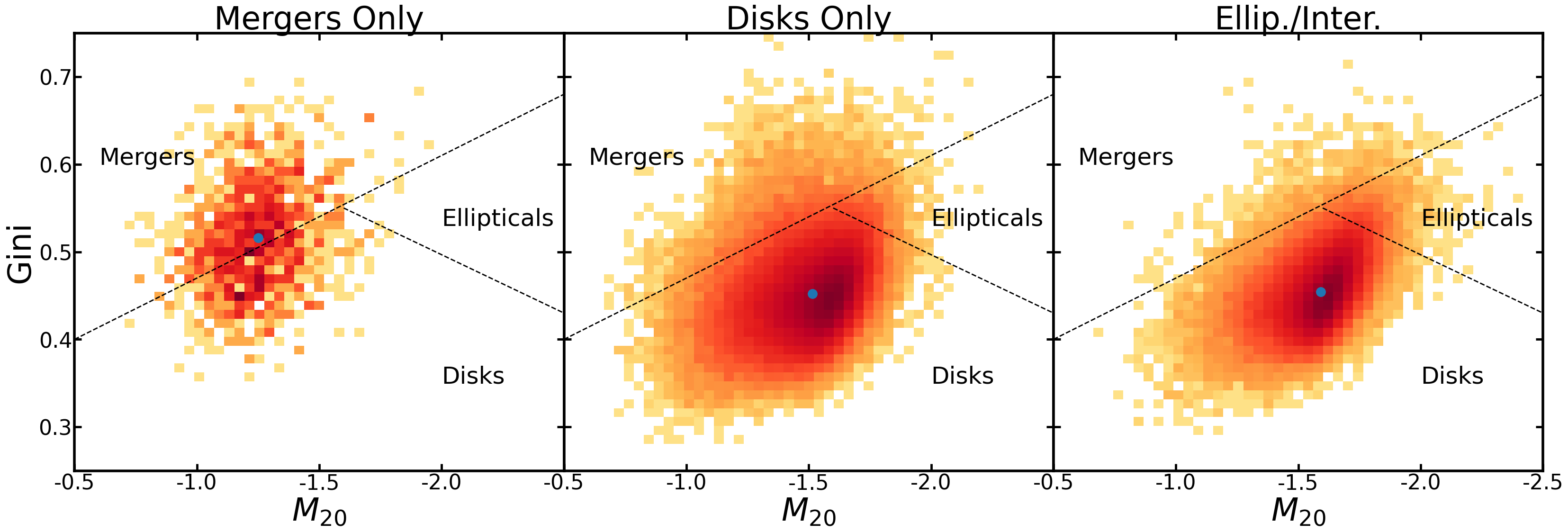}
    \includegraphics[width=\textwidth]{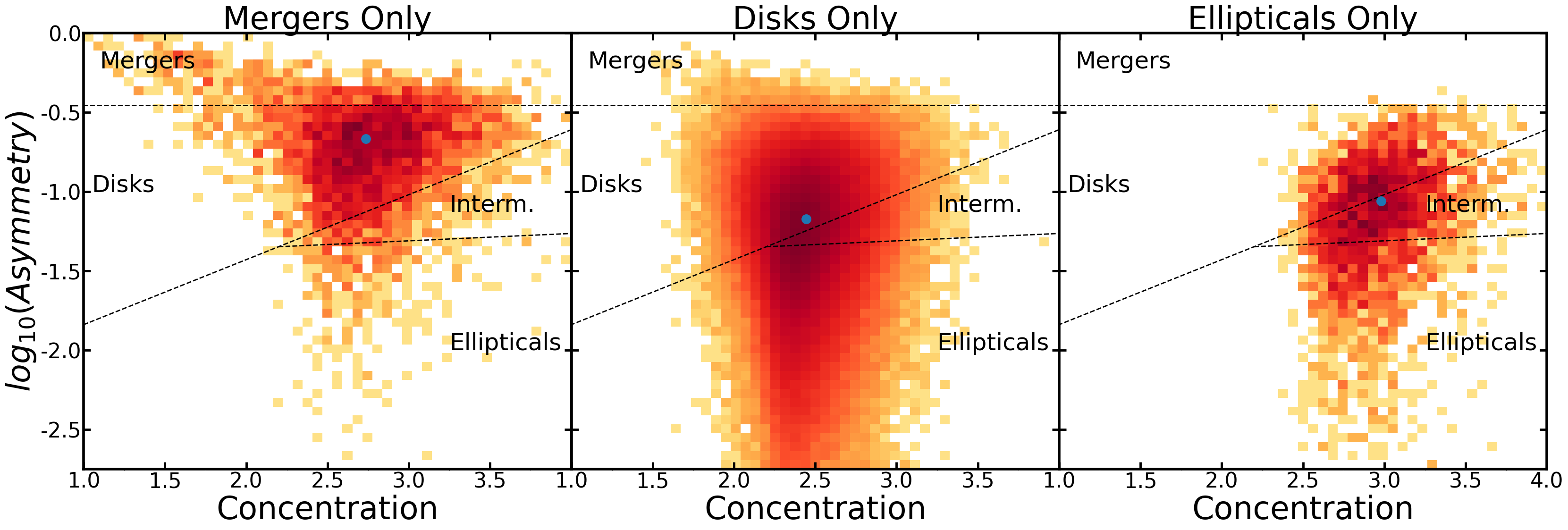}
    \caption{Galaxies classified by the Asymmetry-Concentration bounds, and the Gini-$M_{20}$ bounds plotted on the other axes to compare their classifications. The top row shows the galaxies classified as mergers, disks, and elliptical or intermediate via the Asymmetry-concentration bounds plotted on the Gini-$M_{20}$ plane. The bottom panel shows galaxies classified as mergers, disks, and elliptical via the Gini-$M_{20}$ bounds plotted on the Asymmetry-Concentration plane. The mean and error on the mean of each distribution is marked by the blue marker and error bars.}
    \label{fig:CAS_Gini_M20_comparison}
\end{figure*}

\subsection{Comparing Classification Measures}
\label{sec:classification}

We also analyze how the different classification methods based on morphological fits compare to each other. We do this in two ways. First we divide our population into disks, ellipticals, and mergers based on the CAS boundaries, and the Gini-$M_{20}$ boundaries, and plot them on the opposing axes (i.e. what do galaxies classified as disks by the CAS boundaries look like when plotted on the Gini-$M_{20}$ plane?) which can be seen in Figure \ref{fig:CAS_Gini_M20_comparison}. We also plot our entire sample on the Asymmetry-Concentration and Gini-$M_{20}$ axes, and color it by the average Sersic index within each cell, to analyze how those two classification techniques compare to the Sersic fit.

\subsubsection{Cross-matching CAS and Gini-$M_{20}$ Classifications}

The cross plotting of the CAS classification, and the Gini-$M_{20}$ classification reveals a few important facts. First, we see how few galaxies are classified as mergers based on the $\rm Asymmetry\geq 0.35$ boundary present in the CAS classification. This supports the idea that the Asymmetry bound for mergers is sensitive to redshift, resolution, and wavelength, as the asymmetry bound that is used here was not calibrated for JWST images, nor observations at this redshift. Second, we see that there are fairly few galaxies that are classified as either elliptical or merger based on the Gini-$M_{20}$ classification. This is somewhat expected given the inconclusiveness of the Gini-$M_{20}$ classification discussed in section \ref{sec:non-parametric_measures}. Third, we see that the distributions of galaxies classified as disks, and elliptical based on the CAS statistics have very similar distributions in the Gini-$M_{20}$ plane, which also points to potential issues of the Gini-$M_{20}$ classification.

Despite those issues, there are still valuable takeaways from this analysis. First, despite the similarities between the distributions of the CAS disks and ellipticals plotted on the Gini-$M_{20}$ plane, there is an overall shift up and to the right from the disks to the ellipticals, which can be seen in their mean Gini, and $M_{20}$ values seen in table \ref{table:gini-m20-from-CAS-stats}. While there is noticeable overlap in their distributions, there is a statistically significant difference with a large enough population. Second, even with the smaller sample sizes, the distributions of Gini-$M_{20}$ mergers and ellipticals plotted on the Asymmetry-Concentration plane are noticeably different from the Gini-$M_{20}$ spirals. The Mergers are shifted upwards significantly from the disk population, and the ellipticals are shifted to the right, which do align with the general classification principles of the Asymmetry-Concentration plane.

\subsubsection{Comparing CAS and Gini-$M_{20}$ measurements to Sersic Fits}

Plotting the Asymmetry-Concentration and Gini-$M_{20}$ planes colored by Sersic index also yields interesting results. The Sersic index is not well suited to characterising merging galaxies as irregular and disk type galaxies have a lot of overlap in their Sersic indices, so we focus on the disk vs elliptical/intermediate boundaries. The primary trend in the Asymmetry-Concentration plot is that Sersic index increases with concentration. While this is correlated with the boundary between disks and intermediate/elliptical galaxies, there is also a significant asymmetry component that has effectively no correlation with Sersic index in our data. While this is expected as the concentration and Sersic index both measure how peaked or flat the light profile of a galaxy is, and Sersic fits have no analog for asymmetry, it does show that these two classification metrics rely on different aspects of the galaxy observation, with the Sersic fit relying entirely on the light profile, and the CAS threshold placing much more emphasis on the Asymmetry present in Disk galaxies. 

The Gini-$M_{20}$ plane colored by Sersic index has a trend of increasing Sersic index up and to the right which is roughly perpendicular with the boundary between disks and ellipticals, indicating that they are very correlated. This shows that classifications based on Sersic index, and based on the Gini-$M_{20}$ plot should have similar results. \review{The galaxies that fall in the elliptical portion of the Gini-$M_{20}$ plot have significantly higher Sersic indices ($n=1.2924 \pm 0.0097$) than those in the disk region ($n=0.71072 \pm 0.00040$).} The low number of galaxies classified as elliptical based on the Gini-$M_{20}$ boundaries does align with the low number of galaxies with high Sersic indices in our sample.

While there are some general trends of agreement between the 3 classification techniques examined in this section, there is a great degree of overlap between the distributions classified using each technique when analyzed using the other methods. The populations are distinguishable as the difference in their means are very statistically significant, but the ability for any of these methods to classify individual galaxies from our sample appears very low. This result is not unique to our sample, even when using visual identification as a ground truth, the distributions of disks, elliptical, and irregular galaxies at these redshifts show a great degree of overlap in observational samples like CEERS \citep{CEERS_3}.

\begin{table}
\caption{Gini and $M_{20}$ summary statistics for galaxies classified using the CAS statistics. The locations and errors of the mean points in the upper panel of figure \ref{fig:CAS_Gini_M20_comparison}}
\centering
{\renewcommand{\arraystretch}{1.5}
\begin{tabular}{c|c c c c}
Classification & $M_{20}$ & gini & $\sigma_{M_{20}}$ & $\sigma_{gini}$ \\
\hline
Mergers & -1.2500           & 0.5163     & 0.0053     &     0.0018        \\
Disks & -1.5148           & 0.4520     & 0.0005     &      0.0001       \\
Ellip/S0 & -1.5927          & 0.4544     & 0.0003   &        0.0001       
\end{tabular}
}
\label{table:gini-m20-from-CAS-stats}
\end{table}

\begin{table}
\caption{Concentration and Asymmetry summary statistics for galaxies classified using the gini coefficient and $M_{20}$ statistic. The locations and errors of the mean points in the lower panel of figure \ref{fig:CAS_Gini_M20_comparison}}
\centering
{\renewcommand{\arraystretch}{1.5}
\begin{tabular}{c|c c c c}
Classification & C & A & $\sigma_{C}$ & $\sigma_{A}$ \\
\hline
Mergers & 2.7347           & 0.2143     & 0.0064     &     0.0022        \\
Disks & 2.4441           & 0.0669     & 0.0004     &      0.0001       \\
Ellip & 2.9794           & 0.0867     & 0.0053   &        0.0012       
\end{tabular}
}
\label{table:CAS-from-gm20-stats}
\end{table}

\begin{figure}
    \centering
    \includegraphics[width=1.0\columnwidth]{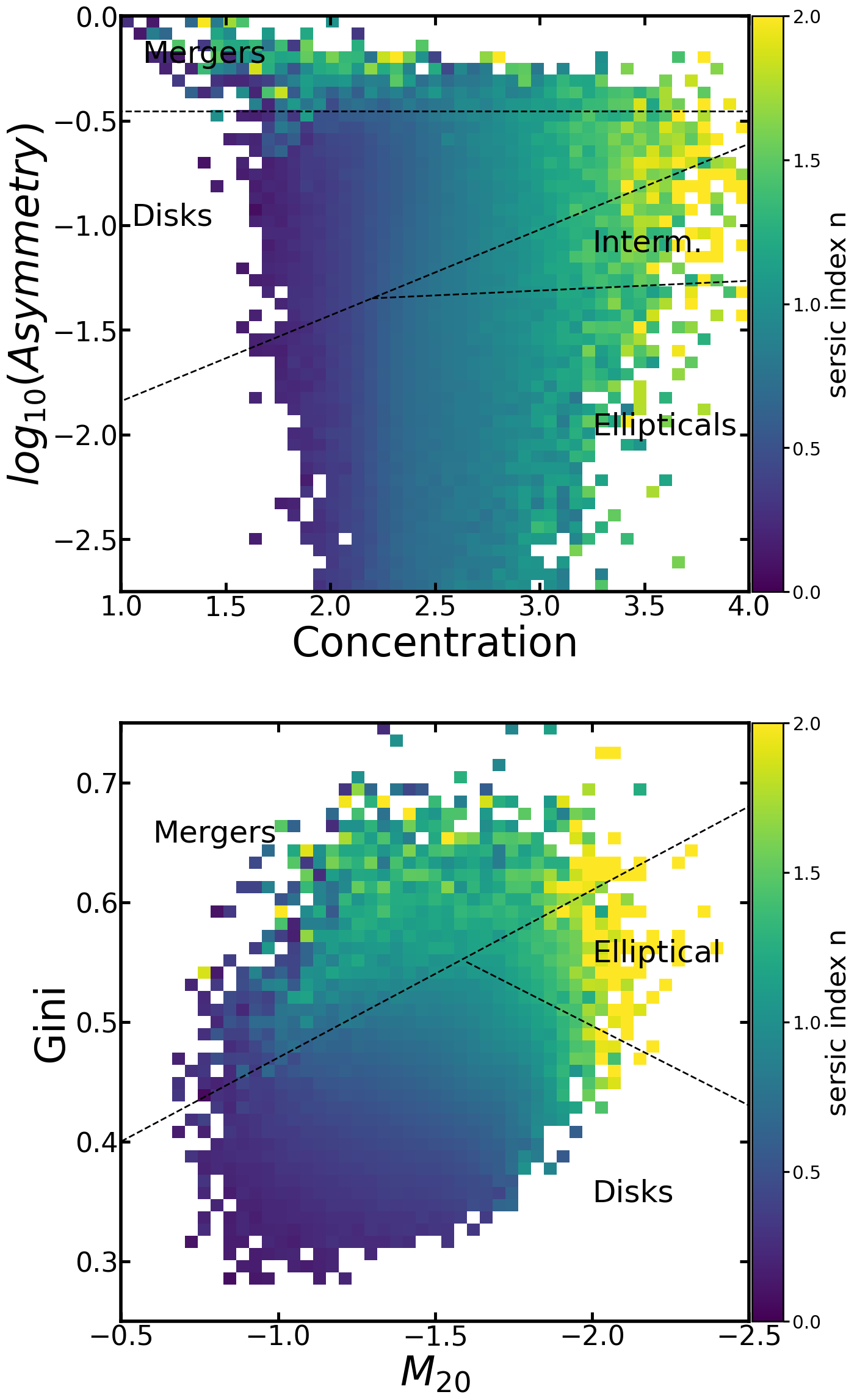}
    \caption{The distribution of galaxies from all redshifts on the Asymmetry-Concentration plane, and Gini-$M_{20}$ plane colored by the average Sersic index in each pixel to investigate how the classification regions of those two plane correlate with classification via Sersic fitting.}
    \label{fig:color_by_n}
\end{figure}

\subsection{AGN Luminosity's potential impact on morphology measurements}
\label{sec:AGN}

We also test the possible influence that light from AGNs could have on galaxy morphology measurements. Works such as \citet{Matthee_2023} investigate the population of AGNs at higher redshifts using JWST, and find there are significantly more than previously expected, and with slightly different properties than expected. Specifically, they use the ${\rm H}\alpha$ broad line emission to identify AGNs, including those that are dim and dust occluded.

\begin{figure}
    \centering
    \includegraphics[width=1.0\columnwidth]{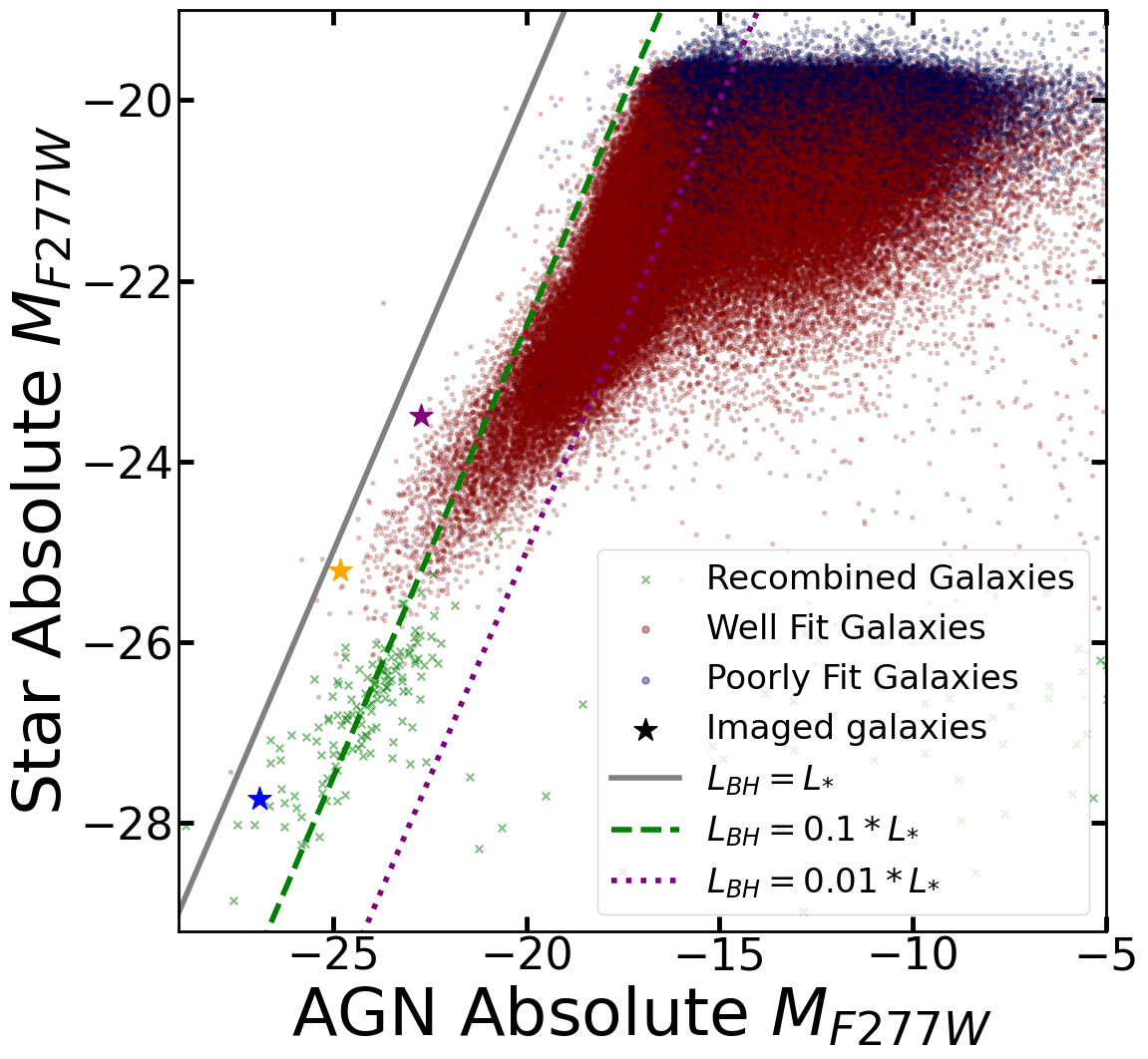}
    \caption{The absolute magnitude of the AGN vs stellar light from each galaxy in the JWST NIRCAM F277W filter. Red and blue points indicate galaxies that were well fit and poorly fit respectively by \texttt{statmorph} before adding the AGN luminosity. The green crosses are galaxies that were recombined with their central regions due to discrepancies with the subhalo catalog. The points indicated by stars are example galaxies we show in greater detail in figure \ref{fig:AGN_images}}
    \label{fig:L_AGN_vs_L_star}
\end{figure}

\begin{figure*}
    \centering
    \includegraphics[width=2.0\columnwidth]{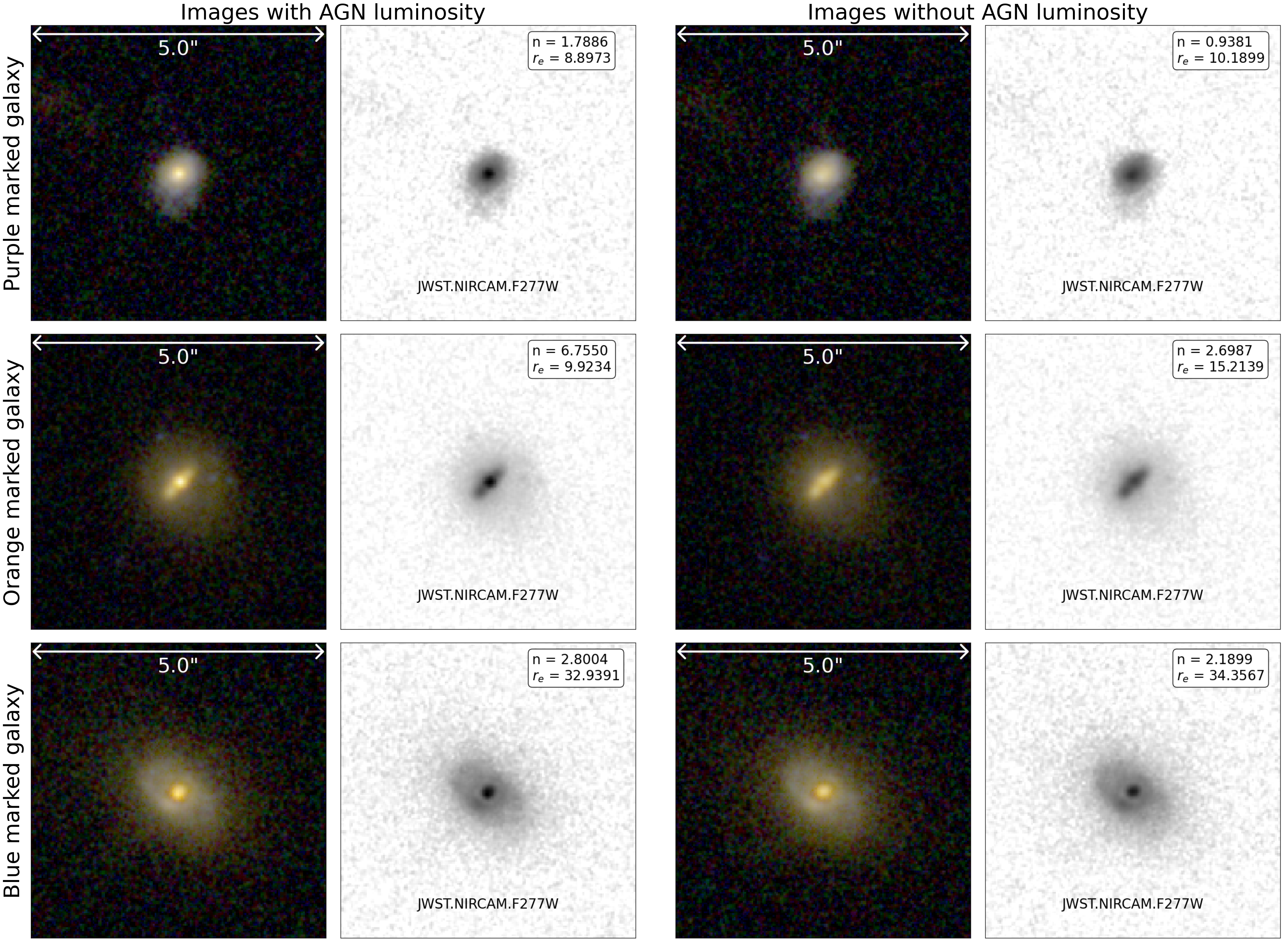}
    \caption{False color, and F277W images of galaxies both with and without the brightness from their AGN. Each row shows a different galaxy, indicated on figure \ref{fig:L_AGN_vs_L_star} by the colored stars. The first row shows the galaxy shown as the purple star, the second row the orange star, and the third row the blue star. The left two images in each row show the galaxy with its AGN, and the right two show it without its AGN. Each row of images shares a color scale to provide proper comparison, but different color scales were used for each row to suit the brightness of each galaxy. Unlike the luminosities show in figure \ref{fig:L_AGN_vs_L_star} these images were created using the entire imaging pipeline, so the effects of dust extinction, background noise, and the telescopes PSF are included.}
    \label{fig:AGN_images}
\end{figure*}

While the number density of these AGNs is low relative to the number density of galaxies overall, their impact on the morphological measures of their host galaxies could be large. We use the AGN SED generation code detailed in \citet{Shen_2020} to find the intrinsic (no dust occlusion) absolute magnitude in the F277W band of AGNs in the redshift $=3$ snapshot and compare those with the intrinsic absolute magnitude of the stars in each galaxy in the same band in Figure \ref{fig:L_AGN_vs_L_star}. While the F277W band adjusted to restframe emission is measuring longer wavelengths than are frequently used in AGN studies, which are largely focused on the restframe UV and X-ray emission, we use this band for two reasons. First, the primary goal of this test is to see if AGN luminosity could impact our measured morphologies which we measure based on the F277W band at redshift $z=3$, so the AGN brightness in that band is most relevant. Second, while our work does not include any spectral analysis, the ${\rm H}\alpha$ emission that is part of the AGN identification criteria used in \citet{Matthee_2023} also falls within the F277W band once adjusted for redshift. Figure \ref{fig:L_AGN_vs_L_star} shows that the vast majority of galaxies have minimal AGN contribution to their overall luminosity in the F277W band compared to the stars. Despite that, there are some galaxies that have significant amounts of AGN luminosity compared to stellar luminosity.

In order to investigate the potential impact those bright AGNs have on their host galaxy's morphology we produce a proof of concept addition to our imaging pipeline that includes the AGN luminosity in our mock observations. Given this proof of concept nature, we only investigate a few galaxies that have significant AGN contributions, and the mock observations of those galaxies can be seen in figure \ref{fig:AGN_images}. Additionally, we perform the same morphological fitting on these images as the rest of our population, and compare the results of each galaxy both with and without the AGN. In each case we see an increase the the Sersic index, and a decrease in $r_{e}$ indicating a more peaked brightness profile. While the number density of galaxies that have bright enough AGNs to influence their morphologies is likely too low to significantly impact the population analysis we perform throughout this paper, the shape of the overall distribution in figure \ref{fig:L_AGN_vs_L_star} indicates the prevalence of AGNs that are bright relative to their host galaxy is much higher for brighter/larger/more massive galaxies. 

During this process, we identified some AGNs that appeared many orders of magnitude brighter than their host galaxies. These are artifacts produced in the subhalo identification process, where the dense central regions of a small proportion of galaxies is identified as a subhalo on its own with just a few star, gas and dark matter particles, and sometimes including the central black hole. This poses an issue for our imaging pipeline, as we use the star gas and black hole particles associated with each subhalo to produce our images. We rectify this for the AGN analysis by identifying subhalos that are within 5 comoving $\rm{kpc} h^{-1}$ of each other and analyzing them together. This selects all of the abnormal subhalos, as they effectively share a position with their host subhalos, without selecting any nearby but distinct subhalos that should be analyzed individually, as 5 comoving$\rm{kpc} h^{-1}$ is approximately 4 long wavelength pixels at redshift 3, so any subhalos this close together would be effectively indistinguishable. While this was not done for the rest of the paper, it is a very small proportion of the overall population and the AGN luminosity dominates the light from these small artificial subhalos, so it has a minimal impact on the analysis presented, as AGN luminosity is not included in that analysis.

\section{Summary and Discussion}
\label{sec:Summary}

In this work we detail the imaging pipeline we created to produce mock NIRCAM observations of galaxies in the \texttt{Astrid} simulation, and use that pipeline to create mock galaxy observations of \review{$\sim 250,000$} well fit galaxies between $z=3$ and $z=6$ and make morphological fits of those galaxies. Using this large population of galaxy images and morphological fits we analyze the population  in a variety of different ways, and reach the following conclusions.
\begin{enumerate}
  \item The galaxy population in \texttt{Astrid} is in rough alignment with the population seen in observations like CEERS \citep{CEERS_3}, and shares many characteristics with other simulated datasets like those created using Illustris TNG \citep{TNG50_CEERS}.
  \item The morphological fits for the \texttt{Astrid} population, and other datasets created from simulations differ from observations in very similar ways, despite different image creation processes indicating potential underlying differences between simulated galaxies and observed galaxies.
  \item While adding in as many instrumentation and physical effects to a mock imaging pipeline should be done, on the population level the impact of each step appears relatively minor, so any significant difference between populations is likely due to factors unrelated to the imaging process.
  \item The Sersic index vs galaxy mass relation, and the mass-size relation both show evolution with redshift, indicating changes in the morphological makeup of galaxies at different masses over time. This evolution is in broad agreement with current work on galaxy population evolution in this time period.
  \item Similar to \citet{TNG50_CEERS} we find that the Sersic effective radius is a good proxy for the half stellar mass radius of galaxies in general, with some caveats at both high redshift and high mass.
  \item  Traditional morphological classification methods based on Sersic index values, and thresholds in the Asymmetry-Concentration, and Gini-$M_{20}$ planes show inconclusive results for our datasets, even when compared to each other.
  \item While the majority of galaxies do not see a meaningful contribution to their luminosity from their central black hole, \review{there is a population that} have a significant portion of their light coming from their AGN, so including it in future mock imaging work will have a significant impact on their measured morphology.
\end{enumerate}

\subsection{Discussion}
\label{subsec:discussion}

The size of the Astrid simulation provides a number of advantages for the creation of a mock observation dataset like this one. It allows for our catalog to include enough galaxies to do statistical analysis of relations that would otherwise not be possible. Additionally, we can examine rare objects that are unlikely to appear in observational surveys like CEERS, or in smaller simulations such as Illustris TNG50. That said, the population of galaxies in our sample is not identical to those in the CEERS catalog, or in the other simulation based catalogs, and understanding how these differences could impact our results is necessary.

\review{The stellar mass function of our sample does differ slightly from CEERS, and while the impact on our results appears minimal, it is worth understanding where this discrepancy comes from. The stellar mass function in Astrid has a roughly uniform shift upwards from the CEERS population, which is likely due to some combination of instrumental factors we did not model, and any selection bias due to our sample existing only at integer redshifts.} 

When comparing the results of morphological fitting, our dataset falls in line with other simulated datasets, but all of those simulated datasets have consistent differences to the CEERS dataset which indicates that there are more fundamental differences between the galaxies in simulations and galaxies in surveys. Specifically, the lack of small galaxies, and high Sersic index galaxies in the simulated datasets is evident, and could have a number of causes. Even state of the art cosmological simulations may lack the mass and spatial resolution to produce these dense objects. Additionally, as we briefly explored adding AGN luminosity could also contribute to addressing this discrepancy. While the population of galaxies with significant AGN brightness after dust occlusion is low, those galaxies would appear smaller, and more concentrated resulting in a higher Sersic index, which would shift them toward the observed population.

We compare the different traditional morphological classification techniques for the galaxies in our sample, and find them relatively inconclusive. This is a common trend amongst higher redshift populations like the CEERS population \citep{CEERS_3}, in that there is overlap between the populations categorized using these classification thresholds. This indicates a potential need for updating the galaxy morphology classification techniques at these higher redshifts, or introducing new techniques such as machine learning based classification using many of the morphological fitting parameters \citep{Rose_23}, or directly from images as in  the work of \citet{Vega-Ferrero_2023}.

Our proof of concept addition of the AGN luminosity to mock galaxy observations has direct applications to this work, but can also be expanded upon to investigate objects such as the "Little Red Dots" that appear significantly more abundant than anticiapated in JWST observations at high redshifts \citep{Matthee_2023}. It is an open question whether these "Little Red Dots" are due in large part due to obscured AGNs as indicated in \citet{Matthee_2023} or if there are other explanations for this population of galaxies like those in \citet{Perez_Gonzalez_2024}. Producing accurate mock galaxy observations with AGN luminosity included could provide insight into this population of objects.

There are a variety of potential future work that could make further use of this dataset. We would like to utilize the size of the catalog to investigate rare objects, analyze cosmic variance, and explore the impact of different cosmic environments on galaxy morphology. Additionally, there is the potential to iterate on the dataset and imaging pipeline used, including adding AGN luminosity to the entire dataset, implementing a more robust dust occlusion code such as \texttt{SKIRT}, and expanding the dataset to include redshifts 1 and 2 now that the \texttt{Astrid} simulation has been run to below redshift 1. Finally, we would like to use this dataset in conjunction with the machine learning based morphological classification techniques that are being developed and/or simulation-based classification techniques like those used in \citet{Dadiani_2023} to further develop galaxy classification at high redshift. The data used and produced in this paper is available from the author on request.

\section*{Acknowledgements}
The \texttt{ASTRID} Simulation was run on the Frontera facility at the Texas Advanced Computing Center.
TDM and RACC acknowledge funding from the NSF AI Institute: Physics of the Future, NSF PHY-2020295, NASA ATP NNX17AK56G, and NASA ATP 80NSSC18K101. TDM acknowledges additional support from  NSF ACI-1614853, NSF AST-1616168, NASA ATP 19-ATP19-0084, and NASA ATP 80NSSC20K0519, and RACC from NSF AST-1909193, SB acknowledges funding supported by NASA-80NSSC22K1897.

%%%%%%%%%%%%%%%%%%%% REFERENCES %%%%%%%%%%%%%%%%%%

% The best way to enter references is to use BibTeX:

\bibliographystyle{mnras}
\bibliography{references} % if your bibtex file is called example.bib

\end{document}